\documentclass[12pt]{article}
\usepackage[utf8]{inputenc}
\usepackage[left=2.5cm,right=2.5cm,top=2.5cm,bottom=3cm]{geometry}
\usepackage{amsmath}
\usepackage{amssymb}
\usepackage{amsfonts}
\usepackage{mathrsfs}
\usepackage{appendix}
\usepackage{mathtools}
\usepackage{tabularray}
\usepackage{bbold}
\usepackage{comment}
\usepackage{cite}
\usepackage{hyperref}
\hypersetup{linktocpage=true,colorlinks=true,citecolor=blue,linkcolor=blue,urlcolor=black}

\usepackage{tikz}
\usetikzlibrary{patterns, quotes, positioning, decorations.markings}
\usetikzlibrary{decorations.pathmorphing}
\tikzset{snake it/.style={decorate, decoration=snake}}
\tikzset{Ibrane/.style={circle, draw=black, fill=black,ultra thick,inner sep=1 pt, minimum size=1 pt,}, c/.default={4pt}}
\title{Higher symmetries of holographic QCD}
\author{Mohammad + Shigeki}
\date{December 2022}
\DeclareMathOperator{\tr}{tr}
\DeclareMathOperator{\Tr}{Tr}
\DeclareMathOperator{\CS}{CS}
\DeclareMathOperator{\dd}{\mathrm{d}}
\DeclareMathOperator{\rme}{\mathrm{e}}
\DeclareMathOperator{\im}{\mathrm{i}}
\DeclareMathOperator{\Z}{\mathbb{Z}}

\makeatletter
\@addtoreset{equation}{section}
\makeatother

\begin{document}
\bibliographystyle{mystyle}
\renewcommand{\baselinestretch}{1.4}
\allowdisplaybreaks[4]
\pagenumbering{gobble}
\vspace{5pt}

\begin{flushright}
\hfill{KUNS-3078}
\end{flushright}

 \begin{center}
 ~\\[10pt]
  \LARGE \bf Generalised anomalies, QCD$_4$, and holography

\vskip 1cm  
\large   
{Mohammad Akhond$^{\dagger\ddagger}$\footnote{\href{mailto:akhond@roma2.infn.it}{akhond@roma2.infn.it}} and}{ Shigeki Sugimoto$^{\dagger *}$\footnote{\href{mailto:sugimoto@gauge.scphys.kyoto-u.ac.jp}{sugimoto@gauge.scphys.kyoto-u.ac.jp}}} \\

\hspace{20pt}

\small  \it
$^{\dagger}$Department of physics, Kyoto University, Kyoto 606-8502, Japan\\
$^{\ddagger}$ INFN – sezione di Roma Tor Vergata \& Dipartimento di Fisica, Via della Ricerca Scientifica, I-00133 Roma, Italy\\
$^*$ Center for Gravitational Physics and Quantum Information,\\
Yukawa Institute of Theoretical Physics, Kyoto University, Kyoto, 606-8502, Japan,\\
$^*$ Kavli Institute for the Physics and Mathematics of the Universe (WPI),\\
The University of Tokyo, Kashiwanoha, Kashiwa 277-8583, Japan\\
\end{center}
\normalsize

\vspace{10pt}
\begin{abstract}
\noindent During the last decade, the notion of an 't Hooft anomaly has been generalised to the case of discrete symmetries. An interesting instance, discussed by Tanizaki, is the mixed anomaly between the discrete axial symmetry and the flavour and baryonic symmetries in massless QCD$_4$. The goal of this note is to provide a derivation of this anomaly from a top-down holographic dual of QCD$_4$. It is found that the topological couplings in the bulk supergravity dual of the D4-D8 system encode Tanizaki's anomaly, once fluctuations around the bulk gauge fields are turned on. A technical challenge for this computation is the difficulty in maintaining gauge invariance of supergravity theories in the presence of D-branes. 
To overcome this issue, a compact formulation of the flux sector of (massive) type IIA supergravity in the presence of D8 branes is presented. A crucial ingredient in the analysis is the need for smearing the D8 branes, in order to impose $\tau$-shift invariance on the fluctuation ansatz.
\end{abstract}
\newpage
\pagenumbering{gobble}
\setcounter{page}{1}
\tableofcontents

\normalsize

\clearpage

\setcounter{footnote}{0}

\pagenumbering{arabic}
\section{Introduction}
\paragraph{}In quantum chromodynamics (QCD), the global 0-form symmetry acting faithfully on the Hilbert space is a quotient of the naive global symmetry $G_\mathrm{naive}$ acting on the quarks, by a subgroup $\mathcal{Z}$ of the centre of the gauge group $G_\mathrm{gauge}$. This means that in the presence of a background for the global symmetry, the structure group $G$ of the total gauge-global bundle takes the form
\begin{equation}\label{gauge/flavour bundle}
   G= \frac{G_\mathrm{gauge}\times G_\mathrm{naive}}{\mathcal{Z}}\;.
\end{equation}
Consequently, when turning on a background for the 0-form symmetry, one needs to include additional data specifying the correlation of the gauge and global bundles. Often this can be achieved by introducing a 2-form gauge field.
\paragraph{}The presence of this additional structure can sometimes lead to new 't Hooft anomalies \cite{Gaiotto:2017tne, Gaiotto:2017yup, Cordova:2019uob, Tanizaki:2018wtg, Sulejmanpasic:2020zfs, Benini:2017dus, Sacchi:2023omn}. The case of interest for the present paper is four-dimensional QCD with $SU(N)$ gauge group, and $N_f$ flavours of massless Dirac fermions in the fundamental representation of $SU(N)$. This theory shall henceforth be referred to as QCD$_4$. The relevant structure group for the global symmetry bundle in this theory is \eqref{symmetry of QCD}. In an interesting paper, Tanizaki demonstrated the presence of a mixed anomaly between the discrete axial symmetry and the flavour symmetry in QCD$_4$ \cite{Tanizaki:2018wtg}.\footnote{Throughout this paper, \emph{flavour} symmetry refers to the vector-like subgroup of the non-abelian chiral symmetry of QCD$_4$, and the abelian component of the vector-like symmetry is referred to as \emph{baryonic} symmetry. This is slightly different from the conventions used in \cite{Tanizaki:2018wtg}.} In the presence of a mass term for the quarks preserving the diagonal component of the chiral symmetry, there is a mixed anomaly between time-reversal and the flavour symmetry when $\theta=\pi$\cite{Cordova:2019uob, Gaiotto:2017tne}. While the main focus of this paper is the massless case, brief comments about the massive case appear in section \ref{Anomaly computation sugra}. The presence of these novel types of discrete anomalies in QCD$_4$ pose a natural question as to their holographic origin. The main motivation for this study is to identify such a holographic interpretation.  
\paragraph{}For field theories with a holographic dual in ten or eleven dimensional supergravity, the anomaly is contained in the bulk topological couplings \cite{Witten:1998qj}. Intuitively, topological terms in supergravity actions are gauge invariant only on closed manifolds. Thus in holographic contexts where the background has a boundary, their gauge invariance is violated by a boundary term, reproducing the anomaly in the dual field theory. Additionally, the topological sector of the bulk supergravity contains BF-type couplings  which encode different global structures of the gauge group in the dual gauge theory, as first noted long ago \cite{Witten:1998wy, Belov:2004ht}. The discovery of more subtle anomalies in field theory, such as those alluded to in the previous paragraphs, has recently prompted renewed interest in more careful investigation of these topological couplings for instance in \cite{Bergman:2020ifi, Bergman:2022otk, Apruzzi:2021phx, Apruzzi:2022rei, Apruzzi:2023uma, Bah:2023ymy, Argurio:2024oym, Bergman:2025isp, Mignosa:2025cpg, Bah:2025vfu, Bergman:2024its, Calvo:2025kjh, Calvo:2025usj}.

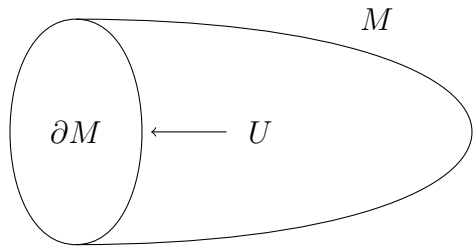
\begin{figure}[!htb]
    \centering
    \begin{tikzpicture}
        \draw (0,-3)to[out=180,in=180,loop,looseness=1](0,0);
        \draw (0,0)to[out=0,in=0,loop,looseness=1](0,-3);
        \draw (0,0)to[out=0,in=0,loop,looseness=6](0,-3);
        \node at (0,-1.5){$\partial M$};
        \node at (4,0){$M$};
        \draw[<-] (1,-1.5)--(2,-1.5);
        \node[label=right:{$U$}] at (2,-1.5){};
    \end{tikzpicture}
    \caption{Schematic of a holographic spacetime $M$. The QFT degrees of freedom live at the boundary $\partial M$, while the holographic coordinate $U$ extends into the bulk spacetime $M$. In this setting it is natural to identify the bulk spacetime of the anomaly TFT with the holographic spacetime $M$.}
    \label{fig:schematic holographic space}
\end{figure}

\paragraph{}Given a background of the form $M\times L$, which is a product of an external manifold $M$, and an internal space $L$, one needs to truncate the bulk topological action, by reducing on the internal space $L$. The resulting theory, is a topological field theory (TFT) living only on $M$, whose action contains the anomalies of the dual field theory, see figure \ref{fig:schematic holographic space} for a schematic representation. The TFT in question also contains the BF-type interactions, which once equipped with suitable boundary conditions, can provide a holographic description of the correlation between gauge and flavour bundles discussed in \eqref{gauge/flavour bundle}.  

\paragraph{}The goal of the present paper is to investigate the topological couplings of the holographic dual of QCD$_4$. This is achieved by working with the top down holographic QCD proposed in \cite{Sakai:2004cn} based on the D4-D8-$\overline{\mathrm{D8}}$ brane system, reviewed in appendix \ref{WSS background}. That the chiral anomaly of QCD$_4$ is encoded in the Chern-Simons couplings of the probe D8 branes, was already demonstrated in \cite{Sakai:2004cn}. The point of the current study is to generalise that discussion by turning on fluctuations for all the bulk supergravity gauge fields, including both RR and NS-NS sectors. This turns out to present some technical challenges as discussed in detail in section \ref{Dual formulation}, with supplementary material tucked away in appendix \ref{SUGRA appendix}. Briefly stated, finding a gauge invariant action for the flux sector of type II supergravities in the presence of D-branes, suitable for studying global properties is subtle. One way to overcome this issue is by employing the \emph{dual} formulation of type IIA supergravity \cite{Bergshoeff:2001pv}. This enables one to write down the TFT that characterizes the anomalies for the system of our interest. Embedded within this TFT are several recently discovered generalised anomalies. These include Tanizaki's mixed axial-flavour and axial-baryon anomalies that show up in the context of massless QCD$_4$ \cite{Tanizaki:2018wtg}. In addition, this TFT contains the so-called anomaly in the space of coupling constants, which is a mixed anomaly between the periodicity of the theta parameter, and the vector-like symmetry of massive QCD. This anomaly was recently investigated in several works, including \cite{Gaiotto:2017tne, Gaiotto:2017yup, Cordova:2019uob}.

\paragraph{}The rest of this paper is organised as follows. Section \ref{sec:field theory} contains a general discussion of anomalies in field theory, and an overview of old and new results on anomalies of QCD$_4$. Section \ref{Dual formulation} contains a discussion of the so-called dual formulation of type IIA supergravity, written in a compact notation which to the authors' best knowledge is original. Appendix \ref{SUGRA appendix} supplements section \ref{Dual formulation} with a discussion of the standard and democratic formulation of type IIA supergravity, and contains several definitions used throughout section \ref{Dual formulation}. In section \ref{Anomaly computation sugra}, the reduction of the bulk topological couplings on the internal space is performed. The effective lower-dimensional TFT is analysed in this section. A general inflow action for QCD$_4$ is derived from the holographic set-up, which is shown to contain the chiral anomaly, as well as the novel types of anomaly discovered recently in \cite{Tanizaki:2018wtg, Cordova:2019uob}. Appendix \ref{WSS background} contains a review of holographic QCD. Section \ref{Conclusions} contains an outlook for future research directions.   
\section{Review: Anomalies in field theory}\label{sec:field theory}
Given a $d$-dimensional QFT on a manifold $\mathcal{M}_d$ with a global symmetry, one can couple the system to a background $\mathcal{B}$.\footnote{The background $\mathcal{B}$ may be that of a discrete or continuous $p$-form symmetry with $p$ generic. In this work, discrete gauge fields are always modeled by embedding into $U(1)$ continuous gauge fields subject to a constraint.} Under background gauge transformation, $\mathcal{B}\to\mathcal{B}^\Lambda$, the partition function need not be invariant and, in general, can pick up a phase
\begin{equation}\label{definition:anomaly}
    Z_{\mathcal{M}_d}[\mathcal{B}^\Lambda]=\exp\left(-2\pi i\int_{\mathcal{M}_d} \mathcal{A}\left(\Lambda,\mathcal{B}\right)\right)Z_{\mathcal{M}_d}[\mathcal{B}]\;.
\end{equation}
When this phase is non-trivial, the theory is said to have an 't Hooft anomaly. One can restore background gauge invariance, by coupling the system to a $(d+1)$-dimensional bulk manifold $\mathcal{N}_{d+1}$, with boundary $\partial \mathcal{N}_{d+1}=\mathcal{M}_d$, supporting degrees of freedom specified by a local Lagrangian $\mathcal{I}_{d+1}(\mathcal{B})$ on $\mathcal{N}_{d+1}$ subject to 
\begin{equation}
    \delta_\Lambda\mathcal{I}_{d+1}(\mathcal{B})=\dd\mathcal{A}(\Lambda,\mathcal{B})
\end{equation}
such that the combined system
\begin{equation}
 \hat{Z}_{\mathcal{N}_{d+1}}[\mathcal{B}]=   Z_{\mathcal{M}_{d}}[\mathcal{B}]\exp\left(2\pi i \int_{\mathcal{N}_{d+1}} \mathcal{I}_{d+1}(\mathcal{B})\right)
\end{equation}
is now invariant under background gauge transformations. This way of thinking gives us a quantitative measure of the anomaly, and the theory specified by the action $\mathcal{S(\mathcal{B})}=2\pi\int\mathcal{I}_{d+1}(\mathcal{B})$ is referred to as the \emph{inflow} action. A theory is said to be anomalous, if its gauge invariant partition function requires specifying a bulk manifold $\mathcal{N}_{d+1}$. Hence the anomaly is trivial if the inflow action is trivial modulo $2\pi\mathbb{Z}$ on closed manifolds. To see this, consider a different choice of a bulk manifold $\mathcal{N}_{d+1}'$. Comparison of the partition function evaluated on $\mathcal{N}_{d+1}$ or $\mathcal{N}_{d+1}'$, shows that they are equivalent if the action evaluated on the closed manifold obtained by gluing $\mathcal{N}_{d+1}$, with $\overline{\mathcal{N}'}_{d+1}$, the orientation reversal of $\mathcal{N}_{d+1}'$ along their common boundary is trivial modulo $2\pi\mathbb{Z}$
\begin{equation}\label{trivial phase}
    \frac{\hat{Z}_{\mathcal{N}_{d+1}}[\mathcal{B}]}{\hat{Z}_{\mathcal{N}_{d+1}'}[\mathcal{B}]}=\exp{\left({2\pi i\int\limits_{\mathcal{N}\cup \overline{\mathcal{N}'}}\mathcal{I}_{d+1}(\mathcal{B})}\right)}=1\;,\quad \text{if}\;\int\limits_{\mathcal{N}\cup \overline{\mathcal{N}'}}\mathcal{I}_{d+1}(\mathcal{B})\in\mathbb{Z}
\end{equation}
\subsection{Anomalies of QCD$_4$}
Given the general definition of anomalies reviewed above, it is interesting to ask about the inflow action for QCD$_4$. As the reader might expect, several studies have already addressed this question. The current subsection serves as a review of existing results, phrased in a notation of our choosing, that makes the holographic match in subsequent sections transparent.
\paragraph{}For the sake of clarity, let us stress once more that, in this paper, QCD$_4$ refers to $SU(N)$ Yang-Mills theory, minimally coupled to $N_f$ flavours of massless Dirac fermions. The latter are composed of $N_f$ left-handed, as well as $N_f$ right-handed two-component Weyl spinors. The Lagrangian for this theory has a global $\left[U(N_f)_L\times U(N_f)_R\right]/\Z_N$ symmetry, where the quotient by $\Z_N$ is due to the fact that a centre gauge transformation must be removed from flavour phase rotations.
The fermion measure in QCD$_4$ suffers an ABJ anomaly which means that the quantum theory has a smaller symmetry group given by 
\begin{equation}\label{symmetry of QCD}
 \frac{SU(N_f)_L\times SU(N_f)_R\times U(1)_V}{\mathbb{Z}_N\times\left(\mathbb{Z}_{N_f}\right)_V}\;.
\end{equation}
The reader is referred to \cite{Tanizaki:2018wtg}, for a detailed analysis and the origin of the quotients in this symmetry group. What follows is a partial survey of known anomalies of QCD$_4$, relevant to the holographic studies in subsequent sections.\footnote{In the presence of a background metric, there are additional anomalies, whose discussion goes beyond the scope of this paper. See however section \ref{Conclusions}, for some comments in this regard.}

\subsubsection{Continuous anomaly}
The oldest, and most appreciated 't Hooft anomaly of QCD$_4$ is the continuous anomaly which shows up when one couples the theory to $SU(N_f)_L\times SU(N_f)_R$ gauge fields denoted by $(A_L,A_R)$. In the presence of this background, the fermion path integral measure picks up a non-trivial Jacobian, which may be canceled by a classical phase in 5d
\begin{equation}\label{chiral anomaly inflow}
    \mathcal{S}[A_L,A_R]=2\pi N\int \left[\CS_5(A_L)-\CS_5(A_R)\right]\;,
\end{equation}
where $\CS_5$ denotes the Chern-Simons five-form.\footnote{The exterior derivative of $\CS_k(\Tilde{A})$ satisfies\begin{equation*}
    \dd \CS_k(\Tilde{A})=\left[\Tr\rme^{\Tilde{F}}\right]_{k+1}\;,
\end{equation*}and the subscript on the right hand side is an instruction to pick the $(k+1)$-form term in the expansion of the exponential. The conventions  used here for the normalization of the gauge field are such that 
\begin{equation*}
    \oint\Tr \Tilde{F}\in\mathbb{Z}\;.
\end{equation*}}
This anomaly is of the original type that 't Hooft had in mind \cite{tHooft:1979rat}, although the name 't Hooft anomaly nowadays refers to a much broader class of anomalies.
\subsubsection{Discrete anomalies}
More subtle anomalies may be detected by careful consideration of the discrete quotients appearing in the symmetry group of QCD \eqref{symmetry of QCD}. In particular, in \cite{Tanizaki:2018wtg}, Tanizaki was able to show that turning on a background for the $\left(\Z_{N_f}\right)_L\times U(N_f)_V/\mathbb{Z}_N$ subgroup of the symmetry group of QCD$_4$ leads to additional discrete anomalies. Here $\left(\Z_{N_f}\right)_L\subset U(1)_L$ is the discrete subgroup of $U(1)_L$ which survives the ABJ anomaly.
Including the dynamical gauge bundle, the total structure group of gauge and flavour symmetry of QCD$_4$ in the presence of this background is\footnote{The quotient by $\mathbb{Z}_N$ on the left hand side corresponds to the identification $(g_c,g_f)\sim(\rme^{\frac{2\pi\im}{N}} g_c,\rme^{\frac{-2\pi\im}{N}}g_f)$, where $g_c\in SU(N)$, and $g_f\in SU(N_f)$. The quotient by $U(1)_*$ on the right hand side corresponds to the identification $(\hat{g}_c,\hat{g}_f)\sim (\rme^{\im\alpha_*}\hat{g}_c,\rme^{-\im\alpha_*}\hat{g}_f)$, where $\hat{g}_c\in U(N)$, $\hat{g}_f\in U(N_f)$, and $\rme^{\im\alpha_*}\in U(1)_*$. Under the natural embedding of $SU(N)\hookrightarrow U(N)$, the two groups are isomorphic.}
\begin{equation}\label{Tanizaki subgroup}
    \frac{SU(N)\times U(N_f)_V}{\Z_N}\times\left(\mathbb{Z}_{N_f}\right)_L\cong\frac{U(N)\times U(N_f)_V}{U(1)_*}\times\left(\Z_{N_f}\right)_L\;.
\end{equation}
Let us denote the gauge field for $U(N)$ by $a$, that for $U(N_f)_V$ by $\Tilde{A}$, and the gauge field for $\left(\mathbb{Z}_{N_f}\right)_L$ by $A_\chi$. The quotient by $U(1)_*$ may be achieved by introducing a background 2-form $B_2$ subject to the constraint\cite{Kapustin:2014gua}
\begin{equation}
    NB_2=\tr(f)\;,
\end{equation}
where $f=\dd a+2\pi \im a^2$ is the field strength for $U(N)$, and assigning the following 1-form gauge transformation properties
\begin{equation}
    B_2\to B_2+\dd \zeta\;,\quad a\to a+\zeta\mathbb{1}_N\;,\quad \Tilde{A}\to \Tilde{A}-\zeta\mathbb{1}_{N_f}\;.
\end{equation}
The kinetic term of the fundamental quarks invariant under the above gauge transformation rules takes the form
\begin{equation}
\Bar{\psi}\gamma^\mu\left(\partial_\mu+a_\mu+\Tilde{A}_\mu+A_\chi\right)P_L\psi+\Bar{\psi}\gamma^\mu\left(\partial_\mu+a_\mu+\Tilde{A}_\mu\right)P_R\psi\;,
\end{equation}
where $P_{L,R}=\frac{1}{2}(1\mp\gamma_5)$. The partition function in the presence of this background is anomalous in the sense of \eqref{definition:anomaly}, and the inflow action is given by
\begin{equation}\label{Tanizaki inflow}
   \mathcal{S}_5\left[A_\chi,\Tilde{A},B_2\right]=\pi N\int A_\chi\wedge\Tr\left[\Tilde{F}^2\right]+2\pi N\int A_\chi\wedge B_2\wedge \Tr \Tilde{F}+\mathcal{O}(A_\chi^2)\;.
\end{equation}
That this expression gives rise to a non-trivial anomaly in the sense of \eqref{trivial phase} follows from the fractional flux quantisation conditions for the discrete fields
\begin{equation}\label{flux quantisation field theory}
    \oint B_2\in\frac{1}{N}\mathbb{Z}\;,\quad \oint A_\chi\in\frac{1}{N_f}\mathbb{Z}
\end{equation}
The inflow action \eqref{Tanizaki inflow} corresponds to a mixed anomaly between the discrete axial symmetry and flavour and baryonic symmetries.

\section{Dual formulation of type IIA fluxes}\label{Dual formulation}
As explained in the introduction, the holographic origin of the sought after anomalies, is the boundary gauge variation of the topological couplings in the bulk. Therefore, the first step in this programme is to find an action governing the topological sector of the bulk theory. This turns out to be a non-trivial problem in the probe brane backgrounds of the type relevant to QCD$_4$. 
 The following section contains a reformulation of the dual formulation of type IIA supergravity \cite{Bergshoeff:2001pv}, restricting attention to the flux sector. In this framework, duality constraints follow as a consequence of the equations of motion, as opposed to being imposed by hand. One can couple the bulk action to D-branes in a gauge invariant manner. In addition, the topological couplings appear manifestly in the Lagrangian. Consequently, this is a suitable framework for studying anomalies of type IIA backgrounds.
 In order to keep the discussion pedagogical, the situation in the absence of D8 branes is considered first. Generalisations to D$p$ branes, in addition to intersecting D$p$-D$q$ brane systems are possible but are not explicitly discussed here. Appendix \ref{SUGRA appendix} contains a supplementary discussion of the standard and democratic formulations of type IIA supergravity and some useful definitions. 
\subsection{Without D-branes}
\paragraph{}The starting point is the action originally proposed in \cite{Bergshoeff:2001pv}, where it was dubbed the \emph{dual} formulation. In the following, only the flux sector of the action is needed, so the metric is assumed to be flat, and the dilaton is set to zero. The bosonic part of the action reads
\begin{equation}\label{dual action G-basis}
\begin{split}
    S&=  -\pi\int \left( \sum_{n=0}^{2}|G_{2n}|^2+|H|^2+ BG_4^2-B^2G_2G_4+\frac{B^3}{3}\left(G_0G_4+G_2^2\right)-\frac{B^4}{4}G_0G_2+\frac{B^5}{20}G_0^2\right)
    \\&- 2\pi\int \left[(G_4-BG_2+\frac{1}{2}B^2G_0)\dd V_5-(G_2-BG_0)\dd V_7+G_0\dd V_9\right]\;,
\end{split}
\end{equation}
where $H=\dd B$, and we work in units where $2\pi l_s=1$. Here, the Kalb-Ramond field $B$ is a 2-form, and the degree of the remaining forms is indicated by a subscript. This action is to be varied with respect to the fluxes $G_{2n}$, the Kalb-Ramond potential $B$, and the Lagrange multipliers $V_{5,7,9}$.
It is convenient to rewrite the above action in a different basis, by introducing fluxes, \footnote{The notation here differs from what was originally used in \cite{Bergshoeff:2001pv}, in particular $\Phi_{\mathrm{here}}=F_{\mathrm{there}}$, in order to reserve $F_{\mathrm{here}}$ to denote the field strength for the D8-brane gauge field.}
\begin{equation}
    \Phi_{2n}=\left(\rme^{-B}G\right)_{2n}\;,\quad n\in\{0,1,2\}\;,
\end{equation}
where the subscript on the right hand side denotes the restriction to the $2n$-form part of the expansion.
In terms of these variables, the action reads
\begin{equation}\label{dual action F-basis}
\begin{split}
    S=&-\pi\int\left[ \sum_{n=0}^{2}\left\lvert\left(e^{B}\Phi\right)_{2n}\right\rvert^2+|H|^2+ \left( B\Phi_4^2+B^2\Phi_2\Phi_4+\frac{B^3}{3}(\Phi_0\Phi_4+\Phi_2^2)+\frac{B^4}{4}\Phi_0\Phi_2+\frac{B^5}{20}\Phi_0^2\right)\right]\\
    &-2\pi\int \left(\Phi_4\dd V_5-\Phi_2\dd V_7+\Phi_0\dd V_9\right)
\end{split}
\end{equation}
It is convenient to define the polyforms
\begin{equation}
    \Phi=\sum_{n=0}^2\Phi_{2n}\;,\quad V=\sum_{n=0}^2 V_{2n+5}\;,
\end{equation}
in terms of which the gauge redundancy of the system may be expressed as
\begin{equation}\label{gauge xfm dual}
    B\to B+\dd\zeta\;,\quad \Phi\to \left(e^{-\dd\zeta}\Phi\right)_<\;,\quad V\to\left[e^{-\dd\zeta}V-\zeta\varphi(-\dd\zeta)\Phi+\dd\Lambda\right]_\geq\;;\quad \Lambda=\sum_{n=0}^2\Lambda_{2n+4}\;,
\end{equation}
where $\varphi(\cdot)$ is defined in \eqref{phi}, and we denote the truncation of a polyform to forms of degree less than (resp. greater than or equal to) five by a subscript $<$ (resp. $\geq$). The gauge parameters, satisfy quantisation conditions
\begin{equation}
    \oint\dd\zeta\in\mathbb{Z}\;,\quad \oint\dd\Lambda\in\mathbb{Z}\;,
\end{equation}
where $\oint$ is used to denote cycle integrals.
It should be stressed that gauge invariance with respect to $\zeta$ is only realised once the constraints from the Lagrange multipliers $V_{5,7,9}$ are imposed. The explicit derivation is presented later, in the discussion leading up to \eqref{gauge xfm dual action}.
The equations of motion following from variation with respect to $\Phi_{0,2,4}$ read
\begin{equation}
    \star\left(e^B\Phi\right)_<=\epsilon_0\left[e^B\left(\Phi+\dd V\right)\right]_>\;,
\end{equation}
where $\epsilon_0$ is defined in \eqref{sign operator}. Acting on this equation with the Hodge star, it may be recast as
\begin{equation}
    \star\left[e^B\left(\Phi+\dd V\right)\right]=\epsilon_0\left[e^B\left(\Phi+\dd V\right)\right]\;,
\end{equation}
which is equivalent to the duality constraint \eqref{duality constraint} in the democratic formulation if one identifies
\begin{equation}\label{duality map}
     \mathcal{G}=e^B\left(\Phi +\dd V\right)\;.
\end{equation}
Similarly, the Bianchi identity for $\dd V$ and the equations of motion for $V$ can be packaged into a single equation
\begin{equation}
    \dd\left(\Phi+\dd V\right)=0\;,
\end{equation}
which can be seen to be equivalent to the Bianchi identity \eqref{Bianchi democratic} in the democratic formulation. In the democratic formulation, the equations of motion are not independent, they follow from combining the Bianchi identity and the duality constraint. As such, we have just demonstrated that the dual formulation agrees with the democratic formulation at the level of the equations of motion.
\paragraph{}The main advantage of working in the $\Phi$-basis of \eqref{dual action F-basis}, as opposed to the $G$-basis of \eqref{dual action G-basis} is that the former is written in terms of quantised fluxes. In particular the flux quantisation conditions read
\begin{equation}\label{flux quantisation w/out D8}
    \oint \dd V\in\mathbb{Z}\;,\quad \oint\Phi\in\mathbb{Z}\;,
\end{equation}
where the first condition is a consequence of the gauge transformation of $V$ with respect to $\Lambda$ in \eqref{gauge xfm dual} and the second condition follows from the path integration over $V$.
\subsubsection{Compact notation}
\paragraph{}The action \eqref{dual action F-basis} may be recast in a more compact form, which will prove its worth when deriving the gauge transformation in the presence of D8 branes. 
It is useful to define a bracket $\langle\cdot ,\cdot\rangle_B$ 
\begin{equation}\label{bracket}
\langle\Phi,\Psi\rangle_B=\int_{M_{10}}\epsilon_0\left(e^B\Phi\right)_<\wedge \left(e^B\Psi\right)_>\;,
\end{equation}
where we have defined two arbitrary polyforms
\begin{equation}
    \Phi=\Phi_0+\Phi_2+\Phi_4\;,\quad \Psi=\Psi_0+\Psi_2+\Psi_4\;.
\end{equation}
We also adopt the convention that a subscript labelled by $<$ (resp. $>$) denotes the truncation to the forms of degree less than (resp. greater than) $5$. In particular we have 
\begin{equation}
\begin{split}
\left(e^B\Phi\right)_<&=\Phi_0+\Phi_2+B\Phi_0+\Phi_4+B\Phi_2+\frac{B^2}{2}\Phi_0\;,\\
\left(e^B\Psi\right)_>&=\left(B+\frac{B^2}{2}+\frac{B^3}{3!}\right)\Psi_4+\left(\frac{B^2}{2}+\frac{B^3}{3!}+\frac{B^4}{4!}\right)\Psi_2+\left(\frac{B^3}{3!}+\frac{B^4}{4!}+\frac{B^5}{5!}\right)\Psi_0\;.
\end{split}
\end{equation}
With the understanding that the integral in \eqref{bracket} only picks up the 10-form part, specifying only one subscript $<$ or $>$ and dropping the other is sufficient
\begin{equation}\label{bracket subscript drop}
\langle\Phi,\Psi\rangle_B=\int_{M_{10}}\epsilon_0\left(e^B\Phi\right)_<\wedge \left(e^B\Psi\right)=\int_{M_{10}}\epsilon_0\left(e^B\Phi\right)\wedge \left(e^B\Psi\right)_>\;.
\end{equation}
Moreover, we have that by definition
\begin{equation}\label{lower}
    \left(e^B\Phi\right)_{<}=e^B\Phi-\left(e^B\Phi\right)_{>}\;.
\end{equation}
These facts can be used to arrive at the following useful identities: 
\begin{enumerate}
    \item The bracket \eqref{bracket} is symmetric:
\begin{equation}\label{symmetric}
    \langle\Phi ,\Psi\rangle_B=\langle\Psi ,\Phi\rangle_B\;.
\end{equation}

\item Shifting the $B$-field by an arbitrary 2-form $\Hat{B}$ while rescaling $\Phi\to\rme^{-\hat{B}}\Phi$, and $\Psi\to\rme^{-\hat{B}}\Psi$ leads to a shift of the bracket by
\begin{equation}\label{B shift bracket identity}
    \left\langle \left(\rme^{-\hat{B}}\Phi\right)_<,\left(\rme^{-\hat{B}}\Psi\right)_<\right\rangle_{B+\Hat{B}}=\langle\Phi ,\Psi\rangle_B-\langle\Phi,\Psi\rangle_{-\Hat{B}}\;.
\end{equation}
\end{enumerate}
The reader is referred to appendix \ref{appendix:proof of bracket identities} for the derivation of these identities.
\paragraph{}The upshot of this somewhat abstract discussion is that we may now recast the action in the dual formulation \eqref{dual action F-basis} as
\begin{equation}\label{dual formulation cpt}
    S=-\pi\int H\wedge\star H+\pi\int\left(\rme^B\Phi\right)_<\wedge\star\left(\rme^B\Phi\right)_< +\pi\left\langle\Phi,\Phi\right\rangle_B+2\pi\int\epsilon_0(\Phi)\wedge \dd V\;.
\end{equation}
Furthermore, the identities \eqref{symmetric}, and \eqref{B shift bracket identity}, together with the flux quantisation conditions \eqref{flux quantisation w/out D8} imply that the gauge variation \eqref{gauge xfm dual} of the action is
\begin{equation}\label{gauge xfm dual action}
\delta S=-\pi\left\langle\Phi,\Phi\right\rangle_{-\dd\zeta}=0\;\text{mod}\;2\pi\mathbb{Z}\;.
\end{equation}
The second equality above follows by assuming that the 10-manifold on which the action \eqref{dual formulation cpt} is formulated is closed, spin, and equipped with additional suitable structure to guarantee the equality.
A sufficient, but not necessary set of conditions is that, the following cycle integrals are always integer quantised
\begin{equation}\label{string structure}
    \oint\exp\left(\dd\zeta\right)\in\mathbb{Z}\;,\quad \frac{1}{2}\oint\Phi_4^2\in\mathbb{Z}\;.
\end{equation}
This can be seen by expanding the gauge variation explicitly
\begin{equation}
    -\pi\left\langle\Phi,\Phi\right\rangle_{-\dd\zeta}=\pi\int \left(\dd\zeta\Phi_4^2+(\dd\zeta)^2\Phi_2\Phi_4+\frac{(\dd\zeta)^3}{3}(\Phi_0\Phi_4+\Phi_2^2)-\frac{(\dd\zeta)^4}{4}\Phi_0\Phi_2+\frac{(\dd\zeta)^5}{20}\Phi_0^2\right)\;.
\end{equation}
Note that the last term is trivial modulo $2\pi\mathbb{Z}$ on closed spin 10-manifolds with vanishing $\hat{A}$-genus, by the index theorem.\footnote{Since the gravitational anomalies are not of interest here, it is assumed that all Stiefel-Whitney classes, and all (fractional) Pontryagin classes of the tangent bundle are trivial.}
On the other hand, the background of interest in this paper, together with the ansatz for the $\Phi_4$ flux \eqref{fluctuation ansatz}, automatically lead to a gauge invariant action.
\subsection{Inclusion of D8 branes}
Recall from the introduction that, ultimately, our goal is to formulate an action for the topological sector of the D4-D8 holographic dual of QCD$_4$.
Our next goal is therefore to include the coupling of D8-branes to the bulk action \eqref{dual action F-basis}. Guided by the discussion of the democratic formulation in appendix \ref{SUGRA appendix}, we seek to establish an action functional for (massive) type IIA supergravity, restricted to flat configurations of the Kalb-Ramond 2-form $H=0$.\footnote{An a posteriori justification for this restriction will be provided in section \ref{Anomaly computation sugra}. 
Essentially it boils down to our choice of the global variant of the gauge group being $SU(N)$, since we wish to consider the theory with fundamental quarks.} With this assumption, one can locally write
\begin{equation}\label{flatness hypothesis}
    B=\dd B_1\;.
\end{equation}
Let $y$ denote the coordinate transverse to the D8 brane, while $A$ is the $U(N_f)$ worldvolume gauge field, and its field strength is denoted by $F=\dd A+2\pi\im A^2$. It is convenient to define a new object $\Omega$ satisfying,\footnote{Note that, in general, one may also include the contribution of the $\Hat{A}$-genus to $\Omega$. This is certainly interesting, and is related to gravitational anomalies. However, it goes beyond the scope of the current work.}
\begin{equation}
    \dd\Omega=\Tr\exp (-F) \delta(y)\dd y\;.
\end{equation}
Given the flatness hypothesis for $B$ \eqref{flatness hypothesis} an explicit local expression for $\Omega$ can be written as
\begin{equation}\label{Omega(A,b)}
   \Omega=\rme^{-B}\Tilde{\Omega}(A,B_1) \;,\qquad \Tilde{\Omega}(A,B_1)=N_f\Theta(y)-\sum_{n\geq 0}(-1)^n\CS_{2n+1}(A-B_1)\delta(y)\dd y\;,
\end{equation}
where $\Theta(y)=\frac{1}{2}\mathrm{sign}(y)$ denotes the step function, and $\CS_{k}$ is the Chern-Simons $k$-form. At first glance, this parameterisation of $\Omega$ seems unnatural - after all $\dd\Omega$ is independent of the $B$-field, while $\Omega$ has a complicated explicit dependence. This parameterisation is chosen so that $\Omega$ transforms covariantly under the $\zeta$ gauge transformations as in \eqref{Kalb-Ramond gauge xfm}. The topological action in the presence of D-branes, can be constructed by requiring its gauge variation to cancel the chiral anomaly residing at the intersection of a D-brane system \cite{Green:1996dd, Cheung:1997az}. 
Ignoring the DBI term, we propose the following as the action for the combined bulk-brane system
\begin{equation}\label{dual formulation compact action}
    \begin{split}
S&=\pi\int\left(e^B\Phi\right)_<\wedge\star\left(e^B\Phi\right)_<+2\pi\int\epsilon_0\left(\Phi+\Omega\right)_<\wedge \left(\dd V-\Omega\right)+\pi\langle\Phi ,\Phi\rangle_B+\pi\int\epsilon_0\Omega_<\wedge\Omega_>\;.
    \end{split}
\end{equation}
Note that this expression contains products of $\Theta(y)$ and $\delta(y)$, which are ill-defined. In the general case, to remove this ambiguity one has to regulate these functions. However, in this paper, we are only interested in a background of the form $M_6\times S^4$, and turn on the gauge field along the $M_6$ direction. In such a situation the would be problematic terms do not appear.
Moreover, in the specific background of interest, we impose $\tau$-shift invariance, as explained in section \ref{Anomaly computation sugra} and appendix \ref{WSS background}, by smearing the D8 brane. The smearing also removes the aforementioned ill-defined terms appearing in this Lagrangian. For the time being, all that is required is to assume the existence of some $\Omega$ with the 1-form gauge transformation prescribed in \eqref{Kalb-Ramond gauge xfm}.
\paragraph{}The action \eqref{dual formulation compact action} is to be varied with respect to RR fluxes $\Phi$, as well as the Lagrange multipliers $V$, while the Kalb-Ramond potential $B$ is treated as a background field. 
The equations of motion for $V$ now impose the constraint
\begin{equation}
    \dd(\Phi+\Omega_<)=0\;,
\end{equation}
Comparison with the modified Bianchi identity for the democratic formulation \eqref{modified bianchi democratic} suggests the map
\begin{equation}
    \rme^{-B}\Tilde{\mathcal{G}}=\Phi+\dd V-\Omega_>\,.
\end{equation}
Path integration over $V$ also implies the modified quantisation condition
\begin{equation}\label{flux quantisation w/ D8}
    \oint_{\Sigma}(\Phi+\Omega_<)\in\mathbb{Z}\;.
\end{equation}
The constituents of the action \eqref{dual formulation compact action} transform under the 1-form gauge symmetry of the Kalb-Ramond field as
\begin{equation}\label{Kalb-Ramond gauge xfm}
\begin{gathered}
    B\to B+\dd \zeta\;,\quad B_1\to B_1+\zeta\;,\quad A\to A+\zeta\mathbb{1}_{N_f}\;,\quad \Omega\to e^{-\dd\zeta}\Omega\;,\\\Phi\to \left(e^{-\dd\zeta}\Phi\right)_<\;,\quad 
     V\to e^{-\dd\zeta}V-\left[\zeta\varphi(-\dd\zeta)(\Phi+\Omega_<)\right]_>\;.
\end{gathered}
\end{equation}
Additionally, there are gauge transformations of the Lagrange multipliers
\begin{equation}
    V\to V+\dd\Lambda\;,
\end{equation}
with $\Lambda$ defined in \eqref{gauge xfm dual}. Finally there is the 0-form gauge redundancy of the D8 brane gauge field
\begin{equation}\label{U(Nf) gauge xfm}
    A\to A^g=gAg^{-1}-\im g\dd g^{-1}\;,\quad \Omega\to\Omega+\dd\alpha(g,A,B_1)\;,\quad V\to V+\alpha(g,A,B_1)_>\;,
\end{equation}
the explicit form of $\alpha(g,A,B_1)$ will not be necessary for the subsequent discussions. The action \eqref{dual formulation compact action}, is not quite gauge invariant with respect to these transformations, rather its gauge transformation cancels the fermion anomaly of D-brane intersections \cite{Green:1996dd, Cheung:1997az}.
\paragraph{}Varying the action \eqref{dual formulation compact action} with respect to $\Phi$ leads to
\begin{equation}\label{eom dual formulation}
   \epsilon_0\star\left(e^B\Phi\right)_<=\left(e^B(\dd V-\Omega_>+\Phi)\right)_>\;,
\end{equation}
which is equivalent to the duality constraint in the democratic formulation provided one identifies
\begin{equation}
    \Tilde{\mathcal{G}}=e^B(\Phi+\dd V-\Omega_>)
\end{equation}
Under the gauge transformations \eqref{Kalb-Ramond gauge xfm}, the kinetic terms are invariant, while the topological couplings transform as
\begin{equation}
    \delta_\zeta S=-\pi\left\langle\Phi+\Omega_<,\Phi+\Omega_<\right\rangle_{-\dd\zeta}=0\;\mathrm{mod}\;2\pi\mathbb{Z}\;,
\end{equation}
on any closed spin 10-manifold, equipped with sufficient additional structure, analogous to those which guarantee \eqref{gauge xfm dual action}. A sufficient set of conditions is obtained by replacing $\Phi_4\to \Phi_4+\Omega_4$ in \eqref{string structure}.
\section{Anomaly matching in holographic QCD}
\label{Anomaly computation sugra}
This section contains a holographic derivation of the anomaly inflow action for QCD$_4$. One way to realise this theory holographically is to consider the D4-D8-$\overline{\text{D8}}$ brane system in type IIA string theory, reviewed in appendix \ref{WSS background}. 
The holographic description is obtained by backreacting the $N$ D4 branes \cite{Witten:1998zw}, while treating the $N_f$ D8-$\overline{\text{D8}}$ pair as probes \cite{Sakai:2004cn}. The resulting background takes the schematic form $\mathbb{R}^{1,3}\times\mathrm{cigar}\times S^4$, where the cigar is a deformed disk parameterised by $U$ and $\tau$ in \eqref{Witten's background}, and $S^4$ denotes the round 4-sphere. In addition to the curved geometry, the background contains a four-form flux
\begin{equation}\label{background flux}
    \int_{S^4} \Phi_4=N\;.
\end{equation}
The reader is referred to appendix \ref{WSS background} for a review of this background, including the explicit metric.
 Let us denote the coordinates on $\mathbb{R}^{1,3}$ by $x^\mu$, with $\mu\in\{0,1,2,3\}$, and use $(y,z)$ defined in \eqref{yz coordinates} to denote the coordinates on the cigar. As discussed in \eqref{D-brane embedding graph} The probe D8 branes are located at $y=0$, and span the remaining directions.
 
 \subsection{Truncation of the topological action on $S^4$}
 Following the philosophy of \cite{Apruzzi:2021phx}, the anomaly theory is obtained by discarding the kinetic terms, and considering fluctuations around the background flux \eqref{background flux}, expanded in a basis of the cohomology of the internal $S^4$. The bulk dynamics of these fluctuations are governed by type IIA supergravity action.
 In section \ref{Dual formulation}, an action functional suitable for precisely such a situation was constructed \eqref{dual formulation compact action}. This action contains gauge invariant kinetic terms, and topological couplings, which are gauge invariant up-to total derivative terms. Hence the anomaly in the holographic setup is encoded in these topological terms - the anomalous phase of the partition function of the dual field theory is reproduced by the anomalous gauge variation of the topological interactions on the $\mathbb{R}^{1,3}\times S^1$ boundary geometry. Discarding the kinetic terms in \eqref{dual formulation compact action} leads one to
\begin{equation}\label{bulk top action}
S_\mathrm{top}=2\pi\int\epsilon_0\left(\Phi+\Omega\right)_<\wedge \left(\dd V-\Omega\right)+\pi\langle\Phi ,\Phi\rangle_B+\pi\int\epsilon_0\Omega_<\wedge\Omega_>\;.
    \end{equation}
Consider the following ansatz for the flux sector
\begin{equation}\label{fluctuation ansatz}
    B_2=b_2=\dd b_1\;,\quad \Phi=\hat{\phi}_0\omega_4+\phi\;,\quad \dd V=\dd\hat{v}\omega_4+\dd v\;,
\end{equation}
where $\omega_4$ is the unit-normalised volume-form of the $S^4$
\begin{equation}
    \int_{S^4}\omega_4=1\;,
\end{equation}
and a compact notation for the fluxes analogous to section \ref{Dual formulation} is used
\begin{equation}
    \phi=\phi_0+\phi_2+\phi_4\;,\quad \hat{v}=\hat{v}_1+\hat{v}_3+\hat{v}_5\;,\quad v=v_5+v_7+v_9\;.
\end{equation}
Note, in particular, that all the dependence on $S^4$ is contained in $\omega_4$. That is, the fluctuations $\{b,\hat{\phi}_0,\phi,\hat{v},v\}$ live purely on the external manifold $M_6=\mathbb{R}^{1,3}\times \mathrm{cigar}$. This ensures invariance under $SO(5)$ isometry of the $S^4$, the significance of this will be emphasised momentarily.

\paragraph{}Substituting the ansatz \eqref{fluctuation ansatz} into the action \eqref{bulk top action}, and integrating over the internal $S^4$, one obtains
\begin{equation}\label{dual formulation reduction}
    S_\mathrm{top}=-2\pi\int_{M_6}\hat{\phi_0}(\dd v_5-\Omega_6)+2\pi\int_{M_6}\epsilon_0(\phi+\Omega)\dd\hat{v}-2\pi\int_{M_6}\hat{\phi}_0e^{b_2}\phi\;.
\end{equation}
Following the discussion in the introduction around \eqref{trivial phase}, our criterion for detecting anomalies, is to evaluate this topological action on a closed manifold, obtained by gluing together $M_6$ with the orientation reversal of some other six-manifold $M_6'$ with an $\hat{\mathbb{R}}^{1,3}\times S^1$ boundary, where $\hat{\mathbb{R}}^{1,3}$ denotes the compactification of $\mathbb{R}^{1,3}$.
The path integral over $v_5$ implies that $\hat{\phi}_0$ is constant and normalised to be $\hat{\phi}_0\in\mathbb{Z}$. In the particular background of interest, the specific value is fixed by \eqref{background flux} to be
\begin{equation}
    \hat{\phi}_0= N\;.
\end{equation}
The equations of motion for $\hat{v}$ read
\begin{equation}
    \dd (\phi+\Omega_<)=0\;,
\end{equation}
whose general solution may be locally written down as
\begin{equation}
   \phi=\dd u+n-\Omega_<\;,
\end{equation}
where $n\in\mathbb{Z}$ is an integration constant and $u=u_1+u_3$. Note that $n$, $u_1$, and $u_3$ are subject to the Dirac quantisation conditions  
\begin{equation}\label{quantisation of c1, c3}
n\in\mathbb{Z}\;,\quad \oint\dd u_1\in\mathbb{Z}\;,\quad \oint\dd u_3\in\mathbb{Z}\;,
\end{equation}
which follow from path integrating over $\hat{v}_5$, $\hat{v}_3$, and $\hat{v}_1$. The topological action obtained after integrating out $\hat{v}_5$, $\hat{v}_3$, and $\hat{v}_1$ reads
\begin{equation}\label{eq:TFT unsmeared}
    S_\mathrm{top}=-2\pi N\int_{M_6} \rme^{b_2}\left(\dd u+n-\Omega \right)\;.
\end{equation}

\paragraph{}
As explained in appendix \ref{WSS background}, Witten's background contains symmetries which are not physical symmetries of QCD$_4$, including the $SO(5)$ isometry of the $S^4$, $U(1)_\tau$ isometry generated by shifts of the $\tau$ coordinate, and a $\mathbb{Z}_2^{(\tau)}$ $\tau$-parity \cite{Brower:2000rp}. In order to extract physical observables of QCD$_4$, rather than some artifact carried by the KK tower, one has to project onto the singlet sector of the unwanted symmetries. This is achieved for the $SO(5)$ symmetry, due to the fluctuation ansatz in \eqref{fluctuation ansatz}, as well as the ansatz for the D8-brane gauge field - no component of the gauge fields is turned on along the $S^4$ direction. 
Although the D8-brane embedding explicitly breaks the isometries of the $\tau$ coordinate, one might still try to restore it by smearing the D8 branes. The reader should note that, at present, this is an assumption of our construction. It is motivated by the fact that the glueball sector of QCD, should be similar to that of pure Yang-Mills theory. It can be shown that the gluons are invariant under $SO(5)$, $U(1)_\tau$ and $\mathbb{Z}_2^{(\tau)}$ and hence the glueballs of pure Yang-Mills should also be invariant under these symmetries as discussed in \cite{Brower:2000rp}. Therefore in the following, projection onto $U(1)_\tau$ and $\mathbb{Z}_2^{(\tau)}$ singlet components of all closed string fields is imposed (see appendix \ref{WSS background}). Imposing $\tau$-parity kills the constant mode $n$ in the $\phi$ fluctuation. After smearing the D8-brane, \eqref{eq:TFT unsmeared} becomes
\begin{equation}\label{topological action}
\begin{split}
    S_\text{top}&=-2\pi N\int_{M_6} e^{b_2}\left(\dd u-\Omega^{(\mathrm{s})}\right),
\end{split}
\end{equation}
where it is understood that the integral picks up only the contribution from the 6-form component of the integrand. Here, $\Omega^{(\mathrm{s})}$ is the smeared version of $\Omega$ as indicated in the superscript, and it is defined as
\begin{equation}
    \Omega^{(\mathrm{s})}=\rme^{-b_2}\Tilde{\Omega}^{(\mathrm{s})}(A-b_1)\;,
\end{equation}
and $\Tilde{\Omega}^{(\mathrm{s})}$ is defined in \eqref{smeared omega}. It is possible to expand the topological action \eqref{topological action} to get the more explicit form
\begin{equation}\label{SymTFT}
  S_\mathrm{top}= - 2\pi N\int_{M_6}\left(b_2\dd u_3+\frac{1}{2}b_2^2\dd u_1-\Tilde{\Omega}_6^{(\mathrm{s})}(A,b_1)\right)
\end{equation}
This action is invariant with respect to the 1-form gauge symmetry
\begin{equation}\label{gauge xfm fluctuations}
    b_2\to b_2+\dd\zeta_1\;,\quad b_1\to b_1+\zeta_1\;,\quad  u\to \left[e^{-\dd\zeta_1}u \right]_{<4}\;,\quad A\to A+\zeta_1\mathbb{1}_{N_f}\;.
\end{equation}
Note however that the fluxes in \eqref{quantisation of c1, c3} are not gauge invariant with respect to the above. This is unavoidable in supergravity \cite{Marolf:2000cb}.
\paragraph{}The terms depending on the D8 brane gauge field in the inflow action read 
\begin{equation}\label{inflow action chiral fields}
    S_\mathrm{top}\supset -2\pi N\int\Tilde{\Omega}_6^{(\mathrm{s})}(A,b_1)=2\pi N\int\left[\CS_5(A_L-b_1)-\CS_5(A_R-b_1)\right]\;.
\end{equation} 
As explained in section \ref{sec:field theory}, the chiral anomaly can be detected by turning on continuous chiral gauge fields, while imposing the fluctuations of the $B$-field to vanish. Setting $b=0$ in \eqref{inflow action chiral fields} yields
\begin{equation}
   \mathcal{S}(A_L,A_R)= 2\pi N\int\left(\CS_5(A_L)-\CS_5(A_R)\right)\;,
\end{equation}
in agreement with the field theory expectation \eqref{chiral anomaly inflow}. This anomaly was already matched holographically in \cite{Sakai:2004cn}.
\subsection{Discrete anomalies}\label{sec: discrete anomaly sugra}
\paragraph{}In the remainder of this section, we aim to see how the discrete anomalies discussed in section \ref{sec:field theory} emerge from the topological action in \eqref{topological action}.
\paragraph{}The first term in the topological action in \eqref{SymTFT} is a BF-term whose role is to specify the global structure of the gauge group of the dual field theory \cite{Witten:1998wy, Bergman:2022otk}. Given that QCD$_4$ contains quarks in the fundamental representation, the global structure of the gauge group must be $SU(N)$, rather than a discrete quotient thereof. The corresponding holographic boundary condition is to let $u_3$ fluctuate on the boundary in \eqref{SymTFT}, while fixing the holonomy of $b_2$ as a $\mathbb{Z}_N$ element. A detailed discussion of the possible topological boundary conditions for this topological action, in the absence of D8-branes appears in \cite{Bertolini:2025wyj}.
For the purpose of our discussion, the relevance of the BF coupling is to set the $b_2$ field to be a $\mathbb{Z}_N$ gauge field. To see that, one can simply consider the theory on $M_4\times S^2$, obtained by compactifying $\mathbb{R}^{1,3}$ to some compact 4-manifold $M_4$ and gluing the cigar with its orientations reversal. The path integral over $u_3$ then imposes
\begin{equation}\label{quantisation of b}
    \dd b_2=0\;,\quad \oint b_2\in\frac{\mathbb{Z}}{N}\;,
\end{equation}
reproducing the desired flux quantisation for the B-field in field theory \eqref{flux quantisation field theory}.
One could view this as an 
a posteriori justification of starting with a bulk supergravity action for $H=0$ configurations in \eqref{dual formulation compact action}.

\paragraph{}It is useful to consider this action formulated on a compact manifold, as discussed around \eqref{inflow action compact}.
This then enables one to combine the $\mathcal{O}(b_2{}^2)$ terms on the D8 brane with the term linear in $\dd u_1$ in the TFT \eqref{SymTFT}, as this combination is gauge invariant
\begin{equation}\label{inflow before imposing stueckelberg}
    2\pi N\int \frac{1}{2}b_2{}^2 \left(\dd u_1+ \frac{1}{2\pi}\Tr (A_L-A_R)\dd\tau\right)+\cdots\;.
\end{equation}
The kinetic term, which was mostly ignored until this point, includes a coupling of the form
\begin{equation}\label{stuckleberg mass}
    S_\mathrm{kin}\supset\pi\int \left\lvert\dd u_1+\frac{1}{2\pi}\Tr( A_L-A_R)\dd \tau\right\rvert^2\;,
\end{equation}
which is a Stueckelberg mass for the axial component of the gauge field. That only the axial component of the gauge field picks up a Stueckelberg mass is a consequence of imposing invariance under $U(1)_\tau$ on the closed string sector fluctuations. At scales well below the Stueckelberg mass, the axial gauge field must satisfy
\begin{equation}\label{minimum stueckelberg}
    \dd u_1+\frac{1}{2\pi}\Tr (A_L-A_R)\dd \tau=0\;.
\end{equation}
One way to argue for this equation is to label $(u_1)_\tau=u_0/(2\pi)$, and define the gauge invariant 1-form
\begin{equation}\label{globally defined 1-form}
    \mathcal{A}=\dd u_0+\Tr(A_L-A_R)
\end{equation}
Since $\mathcal{A}$ is globally defined, it cannot have topologically non-trivial configurations
\begin{equation}
    \oint \dd\mathcal{A}=0\;.
\end{equation}
Moreover, it can be continuously deformed to a trivial configuration $\mathcal{A}\to 0$ that dominates at low energies due to the presence of the mass term \eqref{stuckleberg mass}. To see this, introduce a continuous parameter $t\in[0,1]$, and construct an extrapolating 1-form $\mathcal{A}_t^i$ on each open patch $\mathcal{U}_i$. On double overlaps $\mathcal{U}_i\cap\mathcal{U}_j$, one has
\begin{equation}
    \Tr\left(A_{L,R}^i\right)=\Tr\left(A_{L,R}^j\right)+\dd\alpha_{L,R}^{ij}\;,\quad u_0^i=u_0^j-\left(\alpha_L^{ij}-\alpha_R^{ij}\right)
\end{equation}
Now define a new $U(1)$ gauge field 
\begin{equation}
\hat{A}_t^i=\Tr\left(A_L^i-A_R^i\right)+(t-1)\left[\Tr\left(A_L^i-A_R^i\right)+\dd u_0^i\right]\;,
\end{equation}
where it should be noted that the term inside the square brackets is gauge invariant. 
The globally defined extrapolating 1-form is defined in terms of this $U(1)$ connection as $\mathcal{A}_t^i=\hat{A}_t^i+\dd u_0^i$. As $t\to 0$, $\mathcal{A}_t\to 0$, while $\mathcal{A}_{t=1}$ goes back to the original configuration \eqref{globally defined 1-form}. 
It therefore follows that, at low energies, the axial gauge field must satisfy
\begin{equation}\label{axial gauge field constraints}
    \dd \Tr (A_L-A_R)=0\;,\qquad \oint  \Tr (A_L-A_R)\in \mathbb{Z}\;.
\end{equation}
\paragraph{}It is instructive to rephrase this discussion in terms of the low energy degrees of freedom.
In holographic QCD, the boundary value of $u_1$ is identified with the QCD $\vartheta$ parameter \cite{Witten:1998uka}
\begin{equation}\label{theta parameter}
    \int_\mathrm{cigar}\dd u_1=\int_{S^1}u_1=\mathcal\vartheta\;.
\end{equation}
While the $\eta'$ meson is identified with the integral of the trace of the axial component of the gauge field\footnote{In the unsmeared description, the integral over the $z$ direction of the trace of the D8 brane gauge field is interpreted as the $\eta'$ meson \cite{Sakai:2004cn}
\begin{equation}
    \eta'=\int_\mathrm{cigar}\tr(A)\delta(y)\dd y
\end{equation}}
\begin{equation}
    \frac{1}{2\pi}\int_\mathrm{cigar}\Tr(A_L-A_R)\dd\tau=\eta'
\end{equation}
The combination $(\theta-\eta')$ is natural as it is this particular object which is gauge invariant with respect to the gauge symmetry on the D8 brane.
It was shown in \cite{Sakai:2004cn} that the Stueckelberg action \eqref{stuckleberg mass} contains a mass term for the $\eta'$ meson.

To recover Tanizaki's anomaly, one must restrict to the desired configuration of background fields corresponding to the correct subgroup of the chiral symmetry \eqref{Tanizaki subgroup}
\begin{equation}\label{tanizaki background}
    A_L=\Tilde{A}+A_{\chi}\mathbb{1}_{N_f}\;,\qquad A_R=\Tilde{A}\;.
\end{equation}
The constraints \eqref{axial gauge field constraints}, restricted to this background \eqref{tanizaki background}, are precisely the conditions that characterise $A_\chi$ as a $\mathbb{Z}_{N_f}$ gauge field.
Substituting this configuration into \eqref{inflow action chiral fields}, using \eqref{inflow action expansion}, yields 
\begin{equation}\label{Our Tanizaki action}
\begin{split}
   2\pi N\int\left(\frac{1}{2} A_\chi \wedge \Tr(\Tilde{F}^2)+ A_\chi\wedge b_2\wedge \Tr(\Tilde{F})+\mathcal{O}\left(A_\chi^2\right)\right)\;.
\end{split}
\end{equation}
This is precisely the anomaly written down by Tanizaki in \cite{Tanizaki:2018wtg}. 
\subsection{Coupling constant anomaly of massive QCD$_4$}
Let us now move on to see how the anomaly of massive QCD discussed for instance in \cite{Gaiotto:2017tne,Cordova:2019uob} emerges from the TFT \eqref{SymTFT}. To do so, one must restrict to the appropriate profile for the background gauge fields of the appropriate symmetry group. In particular we must turn on a diagonal component of the flavour symmetry. In the holographic picture this is achieved by setting
\begin{equation}
    A_L=A_R\;.
\end{equation}
With this configuration the topological action \eqref{SymTFT} is reduced to just two terms
\begin{equation}
   2\pi N\int_{M_6}\left(b_2\dd u_3+\frac{1}{2}b_2^2\dd u_1\right)\;.
\end{equation}
In the smeared description used here, one can integrate over the $\tau$ direction to obtain a five-dimensional action\footnote{In order to match the anomaly as presented in \cite{Cordova:2019uob}, one needs to use the dictionary $b_2\rightarrow\frac{1}{N}w_2^{(N)}$, together with \eqref{theta parameter}.}
\begin{equation}\label{Inflow action massive}
    2\pi N\int_{M_5}\left(b_2\dd u_2+\frac{1}{2}b_2^2\dd u_0\right)\;,
\end{equation}
where we have defined
\begin{equation}
    \left(u_2\right)_{MN}=\frac{1}{2\pi}\left(u_3\right)_{MN\tau}\;,\quad u_0=\frac{1}{2\pi}\left(u_1\right)_{\tau}\;.
\end{equation}
The action \eqref{Inflow action massive} matches the inflow action for the anomaly of massive QCD$_4$ reported for instance in \cite{Cordova:2019uob}. 

\section{Conclusions and outlook}\label{Conclusions}

\paragraph{}This work establishes a refinement of the holographic dictionary between large $N$ QCD$_4$, and its top-down supergravity dual. The following contains a summary of our outlook for future research directions.

\paragraph{}Some immediate generalisations are as follows. It is possible to consider turning on the contribution of the $\hat{A}$-genus to the D8 brane Wess-Zumino term. This allows one to study gravitational anomalies. The holographic dual of $SO(N)$, and $Sp(N)$ QCD$_4$ were constructed in \cite{Imoto:2009bf}, and an analysis of the topological sector of those backgrounds could illuminate discrete anomalies of those theories. 

\paragraph{} Similar probe brane backgrounds for QCD$_3$ based on a D3-D7 brane system \cite{Argurio:2020her}, and QCD$_2$ based on a D2-D8 system exist. A holographic study of their generalized anomalies analogous to the present work is desirable. In the case of QCD$_3$, a detailed field theoretic study of the anomalies is available in \cite{Benini:2017dus}.

\paragraph{}One of the main results of this work, is the action for the flux sector of type IIA supergravity discussed in section \ref{Dual formulation}. It would be interesting to consider other IIA supergravity backgrounds with known field theory duals using this dual formulation.

\paragraph{}Recently, a proposal for the holographic dual of $U(N)$ Yang-Mills has been refined in \cite{Bergman:2025isp, Barbar:2025krh}, building on earlier results \cite{Belov:2004ht, Maldacena:2001ss, Witten:1998qj}. It would be interesting to build on this literature and gain an understanding of the holographic dual of $U(N)$ QCD$_4$.
\paragraph{Acknowledgements:}The authors wish to thank Federico Bonetti, Changha Choi, Jaume Gomis, Shinji Hirano, Masazumi Honda, Zohar Komargodski, Carlos Nunez, Andrew O'Bannon, Diego Rodriguez-Gomez, Alejandro Ruiperez, Raffaele Savelli, and Yuya Tanizaki for discussions, comments, and suggestions related to this work. The research of MA is supported by an INFN fellowship under the "Iniziativa Specifica" ST\&FI.
The work of SS is supported by the JSPS KAKENHI (Grant-in-Aid for Scientific Research (B)) grant number JP24K00628 and MEXT KAKENHI (Grant-in-Aid for Transformative Research Areas A “Extreme Universe”) grant number 21H05187.
\begin{appendix}
    \section{Review of top-down holographic QCD}\label{WSS background}
\begin{figure}[!tb]
    \centering
     \begin{tblr}{Q[c,t]Q[c,t]}
     \begin{tikzpicture}[baseline=(current bounding box.north)]
    \draw[thick, decoration={
            markings,
            mark=at position 0.25 with {\arrow{>}},mark=at position 0.75 with {\arrow{>}}
        },
        postaction=decorate
    ](0,0)--(6,0);
    \node[label=above:{$N_f$ D8}]at(2.5,0){};
    \node[label=below:{$N_f$ $\overline{\text{D}8}$}]at(2.5,-2){};
    \draw[thick, decoration={
            markings,
            mark=at position 0.75 with {\arrow{<}},mark=at position 0.25 with {\arrow{<}}
        },
        postaction=decorate
    ](0,-2)--(6,-2);
    \draw[thick, blue] (3,-2)to[out=180,in=180,loop,looseness=.5]node[left]{$N$  $\text{D}4$}(3,0);
    \draw[blue,dashed] (3,-2)to[out=0,in=0,loop,looseness=.5](3,0);

    \draw (0,-2)to[out=180,in=180,loop,looseness=.5](0,0);
    \draw[<-] (-.5,-1.5)to[out=110,in=250,loop,looseness=.5]node[left]{$\tau$}(-.5,-.5);
    \draw[dotted] (0,-2)to[out=0,in=0,loop,looseness=.5](0,0);
    \draw[->](0,-2.5)--(1,-2.5);
    \node[label=left:{$x^{5}$}]at(0,-2.5){};
    \draw (6,-2)to[out=180,in=180,loop,looseness=.5](6,0);
    \draw (6,-2)to[out=0,in=0,loop,looseness=.5](6,0);
    \end{tikzpicture}
  &
    \begin{tikzpicture}[baseline=(current bounding box.north)]
    \draw[<-] (-.5,-1.5)to[out=110,in=250,loop,looseness=.5]node[left]{$\tau$}(-.5,-.5);
 \draw[<-](0,-2.5)--(1,-2.5);
    \node[label=left:{$U$}]at(0,-2.5){};
        \draw (0,-2)to[out=180,in=180,loop,looseness=.5](0,0);
    \draw[dotted] (0,-2)to[out=0,in=0,loop,looseness=.5](0,0);
    \draw[thick, decoration={
            markings,
            mark=at position 0.75 with {\arrow{>}},mark=at position 0.25 with {\arrow{>}}
        },
        postaction=decorate](0,0)to[out=0,in=0,loop,looseness=4](0,-2);
        \node[label=above:{$N_f$ D8}]at(0.5,0){};
    \end{tikzpicture}\\
    (a)&(b)
    \end{tblr}
    \caption{Pictorial representation of the brane system for QCD$_4$, and its holographic dual. (a) Before backreaction the chiral symmetry is realised as the gauge symmetry on the separate stacks of $N_f$ D8, and $\overline{\mathrm{D8}}$ branes. (b) After backreaction of the $N$ D4 branes, the geometry ends at the tip of the cigar, where the $N_f$ D8 and $\overline{\mathrm{D8}}$ branes meet to form a single continuous stack of D8 branes. This phenomenon is interpreted as the holographic manifestation of chiral symmetry breaking.}
    \label{fig:D4-D8 system}
\end{figure}
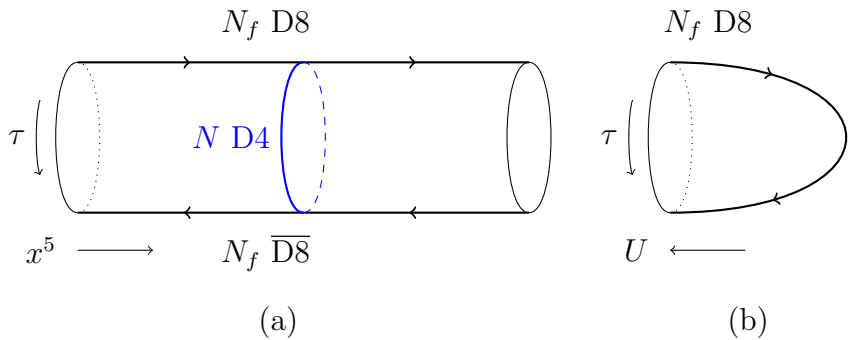

This appendix contains a review of the main features of the top down holographic QCD used in the main text. While it contains sufficient material for the reader to follow the analysis in the main text, more detail can be found in \cite{Sakai:2004cn}.
The type IIA brane system which realises QCD$_4$ can be summarised in the following table
\begin{equation}
    \begin{array}{c}
             \begin{tabular}{c|cccccccccc}
                   & 0&1&2&3&(4)&5&6&7&8&9 \\\hline
                   D4& $\times$&$\times$&$\times$&$\times$&$\times$&&&&&\\
                  D8-$\overline{\text{D}8}$&$\times$&$\times$&$\times$&$\times$&&$\times$&$\times$&$\times$&$\times$&$\times$
             \end{tabular}
             \label{tab:brane scan}
    \end{array}\;.
\end{equation}
The worldvolume theory of the D4 branes is maximally supersymmetric Yang-Mills theory in five dimensions, compactified on a thermal circle along the $x^4$ direction. Due to anti-periodic boundary conditions along $x^4$ for the fermions, a tree level mass is generated for the lowest fermionic Kaluza-Klein mode of the gaugino. Furthermore, a 1-loop mass is generated for the scalar superpartners of the gauge field. 
At high temperatures the theory on the worldvolume is therefore effectively pure Yang-Mills theory in four dimensions. The massless modes of the open strings stretching between the D4 branes and the D8 (resp. $\overline{\text{D}8}$) branes, are left-handed (resp. right-handed) Weyl fermions in the fundamental representation of the $\text{SU}(N)$ gauge group. Thus, the massless modes of this brane system are precisely those of QCD$_4$. Figure \ref{fig:D4-D8 system} depicts a pictorial representation of this brane system.
\paragraph{}In order to obtain a holographic description, one replaces the $N$ D4 branes by Witten's solution \cite{Witten:1998zw}, while treating the D8-$\overline{\text{D}8}$ pair as a probe. This description is valid so long as we keep $N\gg N_f$, i.e. the 't Hooft limit of QCD.
Witten's background, is given by \cite{Witten:1998zw}
\begin{equation}\label{Witten's background}
\begin{gathered}
    ds^2=\left(\frac{U}{R}\right)^\frac{3}{2}\left(\eta_{\mu\nu}dx^\mu dx^\nu+f(U)d\tau^2\right)+\left(\frac{R}{U}\right)^\frac{3}{2}\left(\frac{dU^2}{f(U)}+U^2d\Omega_4^2\right)\;, \\
    e^\phi=g_s\left(\frac{U}{R}\right)^\frac{3}{4}\;,\quad \int_{S^4} \dd C_3=N\;,\quad f(U)=1-\left(\frac{U_{KK}}{U}\right)^3\;,
\end{gathered}
\end{equation}
where $x^\mu$, and $\tau$ parameterise the worldvolume directions of the D4 brane, while $d\Omega_4$ is the line element of $S^4$. The radial holographic coordinate is denoted as $U$, it is bounded from below as $U\geq U_\text{KK}$, $R$ is a constant. The $\tau$ coordinate is periodic, with periodicity
\begin{equation}
    \tau\sim \tau+\frac{2\pi}{M_{KK}}\;,
\end{equation}
where $M_{KK}$ is the Kaluza-Klein mass, related to the parameters of the solution by
\begin{equation}
    M_{KK}=\frac{3}{2}\frac{U_{KK}^{\frac{1}{2}}}{R^{\frac{3}{2}}}\;.
\end{equation}
In what follows, we will always work in units where $M_{KK}=1$.
\subsection{The probe D8 branes}
In order to incorporate massless quarks into the background \eqref{Witten's background}, one introduces $N_f$ D8 branes treated as probes, that is to neglect their backreaction on the geometry. This approximation is valid, provided $N_f$ is an $\mathcal{O}(1)$ number, in the 't Hooft limit.   
\paragraph{}To describe the D8 brane embedding, it is convenient to perform a change of coordinates
\begin{equation}
    U^3=U_{KK}^3\left(1+r^2\right)\;,\;
\end{equation}
and further define
\begin{equation}\label{yz coordinates}
    y=r\cos\tau\;,\quad z=r\sin\tau\;,
\end{equation}
such that $(U,\tau)$ directions are mapped to $(y,z)$. Viewed in the $(y,z)$-plane, the D8-brane embedding takes a rather simple form; the $N_f$ D8 branes are extended along the $z$-direction, and located at $y=0$ as illustrated below
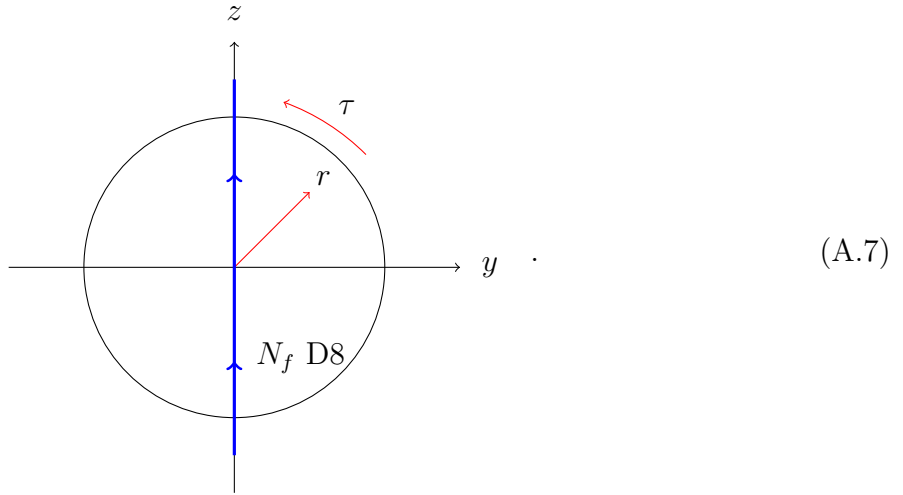
\begin{equation}\label{D-brane embedding graph}
\begin{array}{c}
    \begin{tikzpicture}
         \draw[fill=none](0,0) circle (2) node [black,yshift=-1.5cm] {};
         \draw[red,->](0,0)--(1,1);
         \node[label=above right:{$r$}]at(.8,.8){};
          \draw[red,->] (1.75,1.5) arc (45:70:3);
          \node[label=above:{$\tau$}]at(1.5,1.75){};
         \draw[->](-3,0)--(3,0);
         \draw[->](0,-3)--(0,3);
         \draw[very thick,decoration={
            markings,
            mark=at position 0.75 with {\arrow{>}},mark=at position 0.25 with {\arrow{>}}
        },postaction=decorate,blue](0,-2.5)--(0,2.5);
         \node[label=right:{$N_f$ D8}]at(0,-1.2){};
         \node[label=right:{$y$}]at (3,0){};
         \node[label=above:{$z$}]at (0,3){};
    \end{tikzpicture}
    \end{array}\;.
\end{equation}
A crucial feature of this embedding is that it gives a geometric manifestation of chiral symmetry breaking in QCD. Prior to backreaction, the chiral fermions resided on the separate stacks of $N_f$ D8, and $N_f$ $\overline{\mathrm{D}8}$ branes in figure \ref{fig:D4-D8 system}. However, due to the curvature of the backreacted geometry, the two stacks meet and reconnect at the tip of the cigar - or the origin of the $(y,z)$-plane - to form a single stack of $N_f$ D8 branes. 

The worldvolume gauge field on the D8 brane is identified with the background gauge field for the internal global symmetry of QCD. However, not all components of this gauge field are physical. The background \eqref{Witten's background} has an SO(5) isometry of the internal $S^4$. Clearly such a symmetry is not present in QCD, and thus we should restrict to states invariant under this $\text{SO(5)}$ in order to extract QCD physics. This is ensured by only turning on the D8 brane gauge fields which are $SO(5)$ singlets, namely 
\begin{equation}\label{field strength ansatz}
    F=\frac{1}{2}F_{\mu\nu}\dd x^\mu\wedge \dd x^\nu+F_{\mu z}\dd x^\mu\wedge \dd z\;,\quad \mu,\nu\in\{0,1,2,3\}\;,
\end{equation}
i.e. $F$ has no components along $S^4$ directions.
The holographic interpretation of the background $U(N_f)_L\times U(N_f)_R$ gauge fields is given by the asymptotic value of the D8 brane worldvolume gauge field along the $z$ direction
\begin{equation}\label{background chiral symmetry}
    A_{L\mu}(x^\mu)=\lim_{z\to\infty} A_\mu(x^\mu,z)\;,\quad A_{R\mu}(x^\mu)=\lim_{z\to-\infty} A_\mu(x^\mu,z)\;.
\end{equation}
Writing down a 5d anomaly theory, requires an extension of $A_{L,R}$ to 5d connections. The natural candidate for such an extension is
\begin{equation}\label{chiral gauge fields}
A_\mu(x^\rho,z)=\begin{cases}A^L_{\mu}(x^\rho,z)& z\geq 0\\
A^R_{\mu}(x^\rho,-z)& z\leq 0\\
\end{cases}\;,\quad 
A_z(x^\mu,z)=\begin{cases}
    A^L_{z}(x^\rho,z)&z\geq 0\\
   - A^R_{z}(x^\rho,-z)& z\leq 0
\end{cases}
\end{equation}
In addition to the SO(5) isometry already discussed, the background \eqref{Witten's background} contains further symmetries which are not physical, namely $U(1)_\tau$ isometry generated by shifts of the $\tau$ coordinate, and a $\mathbb{Z}_2^{(\tau)}$ generated by inversion of the $\tau$ coordinate, dubbed $\tau$-parity.

In the presence of D8 branes, $U(1)_\tau$ is explicitly broken by the D8 brane embedding \eqref{D-brane embedding graph}. But since the glueball sector is unaffected by the presence of the massless quarks at least in the leading order in the 't Hooft limit, one might hope to impose invariance under $U(1)_\tau$. This is challenging, if not impossible, because the gauge symmetry on the D8 brane mixes the closed string fields with the open string fields on the D8 brane, as can be seen for instance from \eqref{U(Nf) gauge xfm}. In the following, we use a smeared description of the D8 brane embedding, which allows us to impose invariance under $U(1)_\tau$.

\subsection{Inflow action in the $(y,z)$ coordinates}
Written as an integral over the $z$-direction, the inflow action reads
\begin{equation}\label{inflow action}
   2\pi N\int\limits_{z\in(-\infty,\infty)}\CS_5(A-b_1)=2\pi N\left(\int\limits_{z\in[0,\infty)}\CS_5(A_L-b_1)-\int\limits_{z\in[0,\infty)}\CS_5(A_R-b_1)\right)
\end{equation}
It is useful to consider the compactification of this inflow action, by gluing the cigar with its orientation reversal, and compactifying $\mathbb{R}^{1,3}$ to a compact $M_4$. Assuming the existence of a 6-manifold $N_6$ with boundary $\partial N_6=M_4\times S^1$, the inflow action may be formulated as an integral over $N_6$    
\begin{equation}
    2\pi N\int_{N_6}\frac{1}{3!} \Tr (F-b_2)^3=2\pi N\int_{N_6}\left(\frac{1}{3!}\Tr (F^3)-\frac{1}{2}b_2\Tr(F^2)+\frac{1}{2}b_2{}^2\Tr F+\frac{N_f}{3!}b_2{}^3\right)\;.
\end{equation}
To extract a 5d action on $M_5$, one uses the expansion on the right hand side, together with the flatness of the Kalb-Ramond fluctuations, $\dd b_2=0$, to arrive at
\begin{equation}\label{inflow action compact}
    \int_{M_4\times 
    S^1}\left(\CS_5(A)-b_2\CS_3(A)+\frac{1}{2}b_2{}^2\Tr(A)+\frac{N_f}{3!}b_1b_2{}^2\right)\;.
\end{equation}
In section \ref{sec: discrete anomaly sugra}, the gauge field is restricted to the following configuration
\begin{equation}
   A=A_\chi\mathbb{1}_{N_f}+\Tilde{A}\;,
\end{equation}
with $A_\chi$ only turned on along the left-handed component, while $\Tilde{A}$ is vector-like. The inflow action for this field configuration on $N_6$ reads
\begin{equation}\label{inflow action expansion}
    2\pi N\int_{N_6}\frac{1}{3!}\left(\Tr\left[(\Tilde{F}-b_2)^3\right] +3\dd A_\chi \Tr\left[(\Tilde{F}-b_2)^2\right]+3\left(\dd A_\chi\right)^2\Tr\left(\Tilde{F}-b_2\right)+\left(\dd A_\chi\right)^3 \right)\;.
\end{equation}
From here, one can extract a 5d action, by restricting to the boundary to arrive at \eqref{Our Tanizaki action}. The reader is reminded that the $\mathcal{O}(b_2{}^2)$ term in this action, is to be combined with a term linear in $u_1$, as discussed in \eqref{inflow before imposing stueckelberg}, and subsequent discussions.
\subsection{Symmetry restrictions for the fluctuation ansatz}
\paragraph{} 
As was just argued, in order to extract physical observables of QCD$_4$, rather than some artefact carried by the KK tower, one has to project onto the singlet sector of the unwanted symmetries. This is achieved for the $SO(5)$ symmetry, due to the fluctuation ansatz in \eqref{fluctuation ansatz}, as well as the ansatz for the D8-brane gauge field \eqref{field strength ansatz}. Similarly, projection onto $U(1)_\tau$ and $\mathbb{Z}_2^{(\tau)}$ singlet components of all gauge fields must be imposed. Projection onto $U(1)_\tau$ and $\mathbb{Z}_2^{(\tau)}$ singlet states will play a crucial role in extracting the correct anomaly.

With this in mind, let us postulate that the closed-string sector fluctuations are independent of the $\tau$ coordinate. For instance, the $B$-field fluctuations read
\begin{equation}
    B_{\mu\nu}=b_{\mu\nu}(x^\rho,r)\;,\quad B_{\mu r}=b_{\mu r}(x^\rho,r)\;,\quad B_{\mu\tau}=0\;,
\end{equation}
where following the same conventions as the main text, fluctuations are denoted by small letters. The vanishing of the fluctuations of the $B_{\mu\tau}$ components follows from imposing invariance under $\mathbb{Z}_2^{(\tau)}$. 
\paragraph{}In the absence of D8 branes, one must impose invariance under $\tau$-parity and $\tau$-shift of the remaining supergravity fields $\Phi$, and $V$. This is required for instance, in order to extract the correct glueball spectrum of Yang-Mills theory \cite{Brower:2000rp}.
 
In the presence of D8 brane, one can impose $U(1)_\tau$, and $\mathbb{Z}_2^{(\tau)}$ invariance by smearing the D8 branes. The idea behind smearing \cite{Bigazzi:2005md, Casero:2006pt} is to replace the $\delta$-function appearing in the D8 brane action by a smooth charge distribution. In the specific D4-D8 model considered here, the smeared D8 description was discussed in \cite{Bigazzi:2014qsa, Li:2016gtz}.
Our proposal for the smeared topological action is to replace $\tilde{\Omega}$ in \eqref{Omega(A,b)} with the following smeared version
\begin{equation}\label{smeared omega}
    \Tilde{\Omega}^{(\mathrm{s})}=-\frac{1}{2\pi}\left(\sum_{n\geq 0}(-1)^n\CS_{2n+1}(A_R-b_1)-\sum_{n\geq 0}(-1)^n\CS_{2n+1}(A_L-b_1)\right)\dd\tau
\end{equation}
where the superscript on the left hand side is a reminder that this is a smeared quantity.

\paragraph{}It is particularly relevant to our main analysis to consider the consequence of this symmetry ansatz on the fluctuations of $\Phi_2$. First of all, requiring the fluctuation to be invariant under $U(1)_\tau$ implies that the fluctuations must be independent of the $\tau$-coordinate
\begin{equation}
    \delta\Phi=\phi(x^\mu,r)\;.
\end{equation}
Next, one can further constrain the fluctuations by considering their transformation under $\mathbb{Z}_2^{(\tau)}$.
As discussed in \cite{Imoto:2010ef}, the stringy origin of $\tau$-parity is the combination
\begin{equation}
    P_\tau=(-1)^{F_L}I_{y,9}\;,
\end{equation}
where $(-1)^{F_L}$ acts on RR fields as multiplication by $-1$, and $I_{y,9}$ is an inversion in the $x^9$ and $y$ coordinates. In particular this means that the components of $\phi_2$ have the following transformation
\begin{equation}
   P_\tau:\begin{cases}
       \phi_{MN}\to -\phi_{MN} \\
       \phi_{M\tau}\to \phi_{M\tau}
   \end{cases}\;, 
\end{equation}
where $x^M=(x^\mu,r)$. Therefore, the only allowed non-vanishing components of the fluctuations are
\begin{equation}
    \Phi_2=\phi_{M \tau}\dd x^M\wedge \dd\tau
\end{equation}

After integrating out the Lagrange multiplier $\hat{v}_3$ in \eqref{dual formulation reduction}, and smearing the D8 brane, one may write
\begin{equation}
    \phi_2^{(\mathrm{s})}=\dd u_1+\frac{1}{2\pi}\Tr(A_L-A_R)\wedge \dd\tau
\end{equation}
Note that the $b$-dependence of $\phi_2^{(\mathrm{s})}$ has dropped out and only the axial component of the D8-brane gauge field appears in this expression - this will play a crucial role in our main analysis in section \ref{Anomaly computation sugra}.
\paragraph{}The discussion for $\Phi_4$ is completely analogous. Note that the background value of $\Phi_4=N\omega_4$ is consistent with $\tau$-parity invariance since $I_{y9}$ flips the orientation on $S^4$, and $\Phi_4$ is odd under $(-1)^{F_L}$. For the fluctuations, $\phi_{MNPQ}$ is odd under $\tau$-parity, while $\phi_{MNP\tau}$ is even. Hence, only the latter is allowed to be non-vanishing.
\section{Standard and democratic formulation of type IIA supergravity}\label{SUGRA appendix}
In this section we review the standard and democratic formulations for type IIA supergravity. The two formulations lead to the same equations of motion, though each has certain advantages, depending on the precise application one has in mind. In particular, the democratic formulation makes gauge invariance in the presence of D-branes manifest, while it is not appropriate for studying global or topological properties, since everything is expressed in terms of gauge invariant field strengths. Moreover the duality constraint needs to be imposed after deriving the equations of motion, and so the functional is to be viewed as a pseudo-action. The standard (non-democratic) formulation, can be used to study global properties, as there are topological couplings. However, we are not aware of any successful attempt to write down a fully gauge invariant action that couples the bulk to D-branes using the non-democratic formulation.
\subsection{Standard formulation}
In the standard textbook formulation of supergravity, the flux sector is governed by the action
\begin{equation}\label{non-democrtic action}
\begin{split}
    S&=\pi\int \sum_{n=0}^2G_{2n}\wedge \star G_{2n}-\pi\int H\wedge \star H\\&-\pi\int\left(BG_4^2-B^2G_2G_4+\frac{B^3}{3}\left(G_0G_4+G_2^2\right)-\frac{B^4}{4}G_0G_2+\frac{B^5}{20}G_0^2\right)\;,
\end{split}
\end{equation}
where the dilaton is set to zero, the metric is assumed to be flat,\footnote{Our convention for the Hodge star reads \begin{equation*}
    \star \omega_p =\frac{(-1)^p}{p!(10-p)!}\epsilon_{\mu_1\cdots \mu_{10}}\omega^{\mu_1\cdots\mu_p}\dd x^{\mu_{p+1}}\wedge \cdots\wedge \dd x^{\mu_{10}}
\end{equation*}} and we work in units where $2\pi l_s=1$. In this action, the field strengths are defined in terms of the potentials $C_1$, $C_3$, and $B$
\begin{equation}
    H=\dd B\;,\quad G_4=\dd C_3-HC_1+\frac{B^2}{2}G_0\;,\quad G_2=\dd C_1+B G_0\;,\quad 
\end{equation}
The following Bianchi identities follow from the definition of the field strengths
\begin{equation}\label{sugra bianchi}
    dH=0\;,\qquad \dd G_0=0\;,\qquad \dd G_2=HG_0\;,\qquad dG_4=HG_2\;.
\end{equation}
The action \eqref{non-democrtic action} is invariant under the following gauge transformations
\begin{equation}\label{sugra gauge transformation}
    C_1\to C_1+d\Lambda_0-G_0\zeta_1\;,\quad C_3\to C_3+\dd\Lambda_2-\Lambda_0 H_3-G_0\zeta_1\left(B+\frac{\dd\zeta}{2}\right)\;,\quad B_2\to B_2+ d \zeta \;.
\end{equation}
The equations of motion follow from variation of the action \eqref{non-democrtic action} with respect to the potentials $C$ and $B$
\begin{equation}\label{sugra eom}
    \begin{split}
        d\star G_2&=-H\wedge\star G_4\\
        d\star G_4&=H\wedge G_4\\
        d\star H&=-\frac{1}{2}G_4\wedge G_4+G_2\wedge \star G_4+G_0\wedge\star G_2
    \end{split}
\end{equation}
Finally, we recall that the integer quantised fluxes are the so-called Page charges \cite{Marolf:2000cb}
\begin{equation}\label{sugra flux quantisation}
    \int_{\Sigma_3} H\in\mathbb{Z}\;,\quad \int_{\Sigma_{4}}\left(G_{4}-BG_2+\frac{B^2}{2}G_0\right)\in\mathbb{Z}\;,\quad \int_{\Sigma_2}\left(G_2-BG_0\right)\in\mathbb{Z}\;,\quad G_0\in\mathbb{Z}\;,
\end{equation}
where $\Sigma_{p}$ denotes a closed $p$-dimensional manifold.
\subsection{Democratic formulation}
In the democratic formulation, the flux sector is comprised only of kinetic terms
\begin{equation}\label{democratic action}
    S=\frac{\pi}{2}\int \mathcal{G}\wedge\star \mathcal{G}-\pi\int H\wedge \star H\;,
\end{equation}
where we have introduced 
\begin{equation}
\mathcal{G}=\sum_{n=0}^{5}G_{2n}\;,\quad \mathcal{G}=e^B\left(\dd\left(Ce^{-B}\right)+G_0\right)\;,\quad C=\sum_{n=0}^4C_{2n+1}\quad H=\dd B\;,
\end{equation}
and the integral is understood to pick up only the 10-form part of the integrand.
From these definitions, the Bianchi identity reads
\begin{equation}\label{Bianchi democratic}
    \dd\mathcal{G}=H\wedge\mathcal{G}\;,\qquad\; \dd H=0\;.
\end{equation}
The equations of motion are obtained by varying \eqref{democratic action} with respect to $C$ and $B$ respectively
\begin{equation}
    \dd\star \mathcal{G}=-H\star \mathcal{G}\;,\quad \dd \star H=\frac{1}{2}\mathcal{G}\wedge \star\mathcal{G}
\end{equation}
The action \eqref{democratic action} is invariant under the following gauge symmetries
\begin{equation}\label{gauge transformation without d-branes democratic}
    B\to B+\dd\zeta\;,\qquad C\to C-G_0\zeta \rme^{B}\varphi(\dd\zeta)+e^B\dd\left(\Lambda e^{-B}\right) \;;\quad \Lambda=\sum_{n=0}^4\Lambda_{2n}\;,
\end{equation}
where we introduced the function \cite{Sugimoto:20XX}
\begin{equation}\label{phi}
    \varphi(x)=\frac{e^x-1}{x}=\sum_{n\geq 0}\frac{x^n}{(n+1)!}=1+\frac{x}{2!}+\frac{x^2}{3!}+\cdots\;.
\end{equation}
In addition to the equations of motion, one must impose the duality constraints
\begin{equation}\label{duality constraint}
    \mathcal{G}=\epsilon_0\star \mathcal{G}
\end{equation}
after deriving the equations of motion.
In the above $\epsilon_0$ is the sign operator defined via
\begin{equation}\label{sign operator}
    \epsilon_i A_n=\begin{cases}
        -A_n & n=i\;\text{mod}\;4\\
        +A_n&\text{Otherwise}
    \end{cases}
\end{equation}
where $A_n$ is any $n$-form.
The following useful identities follow easily from the above definition, firstly note that the index of the sign operator is well defined mod 4
\begin{equation}
    \epsilon_i=\epsilon_{i\pm4}\;.
\end{equation}
Given any $m$-form $B_m$, we have
\begin{equation}
    \epsilon_i\left(A_n\wedge B_m\right)=A_n\wedge\epsilon_{i-n}B_m=\epsilon_{i-m}A_n\wedge B_m
\end{equation}
and
\begin{equation}
   \epsilon_i\dd A_n=\dd\left(\epsilon_{i-1}A_n\right) 
\end{equation}
\paragraph{Flux quantisation} The fluxes are subject to Dirac quantisation conditions which read
\begin{equation}
    \int H\in\mathbb{Z}\;,\quad \int e^{-B}\mathcal{G}\in\mathbb{Z}
\end{equation}
\subsection{Inclusion of D8 branes}
In the presence of D-branes, the action needs to be modified to include the bulk-brane couplings. One elegant way to determine these couplings is to use the constraints from cancellation of chiral anomaly on the worldvolume of certain D-brane intersections \cite{Green:1996dd, Cheung:1997az, Schwarz:2001sf}. 
\paragraph{}One advantage of the democratic formulation is that it allows for including the coupling to D-branes. Consider, for instance, the coupling to a D8 brane
\begin{equation}
    S_\mathrm{D8}=S_\mathrm{DBI}+S_{\CS}\;.
\end{equation}
The first term is the usual DBI action, which will not play a crucial role in our analysis. The second term is the CS coupling, which can be determined using anomaly inflow arguments \cite{Green:1996dd, Cheung:1997az}.
Denoting the coordinate transverse to the D8 brane by $y$, it is useful to define
\begin{equation}
    Y=\Tr \exp\left(F\right)\delta(y)\dd y\;,
\end{equation}
where $F=\dd A+\im A^2$ is the field strength of the D8 brane gauge field.
Restricting to field configurations where $H=0$, one may locally write
\begin{equation}
    B_2=\dd B_1\;.
\end{equation}
It is then possible to write a local expression 
\begin{equation}
    Y=-\epsilon_1\dd\Omega\;,
\end{equation}
with $\Omega$ defined in \eqref{Omega(A,b)}. Under the gauge symmetry $A\to gAg^{-1}-\im g\dd g^{-1}$ on the D8 brane, $\Omega$ transforms as a total derivative
\begin{equation}
    \delta\Omega=\dd\alpha(g,A,B_1)\;.
\end{equation}
 Crucially, the bulk RR fields have a non-trivial transformation under this symmetry \cite{Green:1996dd, Cheung:1997az, Schwarz:2001sf}
\begin{equation}
  \delta\left(Ce^{-B}\right)=\alpha(g,A,B_1);.
\end{equation}
This is the requirement for cancellation of the chiral anomaly \cite{Green:1996dd, Cheung:1997az}. Accordingly, the kinetic terms need to be modified in order to preserve gauge invariance. This is achieved by replacing
\begin{equation}
    \mathcal{G}\to \tilde{\mathcal{G}}=e^B\left(\dd\left(Ce^{-B}\right)-\Omega+G_0\right)\;.
\end{equation}
These fluxes satisfy a modified Bianchi identity
\begin{equation}\label{modified bianchi democratic}
    \dd\left(\rme^{-B}\tilde{\mathcal{G}}\right)=-\dd\Omega
\end{equation}
Invariance under the Kalb-Ramond gauge symmetry $B\to B+\dd\zeta$ is ensured by postulating that the RR potentials transform as
\begin{equation}\label{dzCe^-B}
    Ce^{-B}\to Ce^{-(B+\dd\zeta)}-G_0\zeta \varphi(-\dd\zeta)\;.
\end{equation}
The action for the bulk-brane system then takes the following form
\begin{equation}
     S=\frac{\pi}{2}\int_{M_{10}} \Tilde{\mathcal{G}}\wedge\star \Tilde{\mathcal{G}}-\pi\int_{M_{10}} H\wedge \star H+\pi\int_{M_{11}}e^{-B}\mathcal{G}\wedge Y +S_\mathrm{DBI}\;,
\end{equation}
where we have written the D8 brane CS term as an integral over an 11d manifold $M_{11}$ that bounds $M_{10}$, namely $\partial M_{11}=M_{10}$. Gauge invariance of the kinetic terms is manifest, since $\Tilde{\mathcal{G}}$ is invariant by construction. There are two equivalent 10d expressions for the D8 brane CS coupling, which differ by total derivatives
\begin{equation}\label{CS D8 democratic}
 S_{\CS}^\mathrm{D8}=\pi   \int_{M_{10}}\left(Ce^{-B}\wedge Y+G_0\Omega_{10}\right)=\pi\int_{M_{10}} \epsilon_{03}\left(e^B\Omega\right)\wedge \Tilde{\mathcal{G}}\,
\end{equation}
where $\epsilon_{03}=\epsilon_0\epsilon_3$.
The gauge variation with respect to the D8 brane gauge field is engineered to cancel the fermion anomaly \cite{Green:1996dd, Cheung:1997az}.
\section{Proof of bracket identities \eqref{symmetric}, and \eqref{B shift bracket identity}}\label{appendix:proof of bracket identities}
This appendix contains detailed derivations of the bracket identities \eqref{symmetric}, and \eqref{B shift bracket identity} stated in the main text.
The first identity \eqref{symmetric} follows from inserting \eqref{lower} in the middle expression in \eqref{bracket subscript drop}, so that the left hand side of \eqref{symmetric} becomes
\begin{equation}\label{proof symmetric}
   \begin{split} \langle\Phi ,\Psi\rangle_B&=\int\epsilon_0\left[\rme^B\Phi-\left(\rme^B\Phi\right)_>\right]\wedge\rme^B\Psi\\
   &=\int \epsilon_0\left(\Phi\right)\wedge\Psi-\int \epsilon_0\left(\rme^B\Phi\right)_>\wedge\left(\rme^B\Psi\right)_< 
   \end{split}\;,
\end{equation}
where we used the fact that $\epsilon_0\left(\rme^B\Phi\right)=\rme^{-B}\epsilon_0\left(\Phi\right)$ to arrive at the first term in the second line. At this point, one notes that the first term in the second line of \eqref{proof symmetric}, does not contribute to the integral, as it does not contain a 10-form component. The remaining term in the second line of \eqref{proof symmetric} can be rearranged using the definition of the $\epsilon_0$ operator, as the reader can readily confirm by expanding in components, arriving at
\begin{equation}
    \langle\Phi ,\Psi\rangle_B=-\int \epsilon_2\left(\rme^B\Psi\right)_<\wedge \left(\rme^B\Phi\right)_> =+\int \epsilon_0\left(\rme^B\Psi\right)_<\wedge \left(\rme^B\Phi\right)_>= \langle\Psi ,\Phi\rangle_B\;.
\end{equation}
This establishes a proof of the identity \eqref{symmetric}.

To prove \eqref{B shift bracket identity} one inserts the definition of the bracket \eqref{bracket} into the left hand side of \eqref{B shift bracket identity}
\begin{equation}\label{bracket shift step1}
\begin{split}
    \left\langle \left(\rme^{-\hat{B}}\Phi\right)_<,\left(\rme^{-\hat{B}}\Psi\right)_<\right\rangle_{B+\Hat{B}}&=\int\epsilon_0\left(\rme^{B+\hat{B}}\left(\rme^{-\hat{B}}\Phi\right)_<\right)_<\wedge \left(\rme^{B+\hat{B}}\left(\rme^{-\hat{B}}\Psi\right)_<\right)_>\;\\
    &=\int\epsilon_0\left(\rme^{B}\Phi\right)_<\wedge \left(\rme^{B+\hat{B}}\left(\rme^{-\hat{B}}\Psi\right)_<\right)\;.
    \end{split}
\end{equation}
In the second line we used the fact that the integral only picks up the 10-form component of the integrand to drop the subscript $>$, and used the identity
\begin{equation}
\left(\rme^{B+\hat{B}}\left(\rme^{-\hat{B}}\Phi\right)_<\right)_<=\left(\rme^B\Phi\right)_<\;.
\end{equation}
One should be careful, that a similar simplification cannot occur in the expression containing the $\Psi$ field, that is
\begin{equation}
\left(\rme^{B+\hat{B}}\left(\rme^{-\hat{B}}\Psi\right)_<\right)_>\neq\left(\rme^B\Psi\right)_>\;,
\end{equation}
because the factor $\rme^{\hat{B}}$ contains forms up to degree 10, while $\rme^{-\hat{B}}$ is truncated as instructed by the subscript $<$. To proceed further, one uses the analogue of \eqref{lower} for $\Psi$ into the second line of \eqref{bracket shift step1}
\begin{equation}\label{bracket shift step2}
\begin{split}
    \left\langle \left(\rme^{-\hat{B}}\Phi\right)_<,\left(\rme^{-\hat{B}}\Psi\right)_<\right\rangle_{B+\Hat{B}}&=\int\epsilon_0\left(\rme^{B}\Phi\right)_<\wedge \rme^{B+\hat{B}}\left[\rme^{-\hat{B}}\Psi-\left(\rme^{-\hat{B}}\Psi\right)_>\right]\\
    &=\int\epsilon_0\left(\rme^B\Phi\right)_<\wedge\left(\rme^B\Psi\right)-\int \epsilon_0\left(    \rme^B\Phi\right)\rme^{B+\hat{B}}\left(\rme^{-\hat{B}}\Psi\right)_>\;,
    \end{split}
\end{equation}
where once again a subscript $<$ is dropped in the second term in the second line, exploiting the fact that the integral only picks up the 10-form component of the integrand. Using the identity, $\epsilon_0\left(\rme^B\Phi\right)\rme^{B+\hat{B}}=\epsilon_0\left(\rme^{-B-\hat{B}}\rme^B\Phi\right)=\epsilon_0\left(\rme^{-\hat{B}}\Phi\right)$, one finds that the right hand side of \eqref{bracket shift step2} is equivalent to the right hand side of \eqref{B shift bracket identity}.
\end{appendix}
\bibliography{ref}
\end{document}